\documentclass[aps,reprint,twocolumn,pre,superscriptaddress,10pt]{revtex4}
\usepackage{amsmath}
\usepackage[pdftex]{graphicx}
\usepackage{amsmath}
\usepackage{amssymb}
\usepackage[usenames,dvipsnames]{color}
\newcommand{\es}{\enspace}
\newcommand{\beq}{\begin{equation}}
\newcommand{\eeq}{\end{equation}}
\newcommand{\beqa}{\begin{eqnarray}}
\newcommand{\eeqa}{\end{eqnarray}}

\newcommand{\Ein}[1]{\langle#1\rangle}
\newcommand{\KL}{{\rm KL}}
\newcommand{\tij}{{\tau_{(i,j)}}}
\newcommand{\dti}{{\Delta\tau_i}}
\newcommand{\dtij}{{\Delta\tau_{(i,j)}}}
\newcommand{\fa}{{f_{\rm a}}}
\newcommand{\Pa}{{\Pi_{\rm a}}}
\newcommand{\fl}{{f_\ell}}
\newcommand{\Pl}{{\Pi_\ell}}

\begin{document}
%===== Title ================================
\title{How memory generates heterogeneous dynamics in temporal networks}
\author{Christian L. Vestergaard}
\email{cvestergaard@gmail.com}
\affiliation{Aix Marseille Universit\'{e}, Universit\'e de Toulon, CNRS, CPT, UMR 7332, 13288 Marseille, France}
\author{Mathieu G\'enois}
\affiliation{Aix Marseille Universit\'{e}, Universit\'e de Toulon, CNRS, CPT, UMR 7332, 13288 Marseille, France}
\author{Alain Barrat}
\affiliation{Aix Marseille Universit\'{e}, Universit\'e de Toulon, CNRS, CPT, UMR 7332, 13288 Marseille, France}
\affiliation{Data Science Laboratory, ISI Foundation, Torino, Italy}

\date{\today}

\begin{abstract}
Empirical temporal networks display strong heterogeneities in their dynamics, which
profoundly affect processes taking place on these networks, such as rumor and epidemic spreading.
Despite the recent wealth of data on temporal networks, little work has been devoted
to the understanding of how such heterogeneities can emerge from microscopic mechanisms at the level of nodes and links.
Here we show that long-term memory effects are present in the creation and disappearance of links in empirical networks.
We thus consider a simple generative modeling framework for temporal networks able to incorporate these memory mechanisms.
This allows us to study separately the role of each of these mechanisms in the emergence of heterogeneous network dynamics.
In particular, we show analytically and numerically how heterogeneous distributions of contact durations, of inter-contact durations and of numbers of contacts per link emerge.
We also study the individual effect of heterogeneities on dynamical processes, such as the paradigmatic Susceptible-Infected epidemic spreading model.
Our results confirm in particular the crucial role of the distributions of inter-contact durations and of the numbers of contacts per link.
\end{abstract}

\maketitle
%===== Body: ================================
\section{Introduction.}
The availability of large temporally resolved datasets of human communication~\cite{Eckmann2004b,Vazquez2006,Onnela2007,Rybski2009,Karsai2012,Karsai2014} and mobility~\cite{Barrat2004,Gonzalez:2008,Gautreau2009,Balcan:2009a,Rosvall2013}, as well as recent technological advances allowing the recording of physical proximity~\cite{Salathe2010,Stopczynski2014} and face-to-face contacts in social groups~\cite{Cattuto2010,Barrat2013b}, has made it possible to study the dynamics of complex networks.
These systems show heterogeneous dynamics, which crucially affect dynamical processes taking place on the networks, such as rumor or epidemic spreading~\cite{Vazquez2007,Miritello2011a,Karsai2011b,Panisson2012,Gauvin2013,Holme2013a,Karsai2014}.
While the role of various individual mechanisms in the emergence of the heterogeneous topology of slowly varying networks has been well studied~\cite{Ghoshal2013}, less work
has been devoted to the development of models for temporal networks. It is still unclear how their heterogeneous dynamics emerge from microscopic interaction mechanisms at the level of single nodes and links.
There is in particular a need for generative models containing a minimal number of plausible microscopic mechanisms,
both to understand how temporal heterogeneities emerge and
to serve as simple yet realistic paradigms of temporal networks for the study of how dynamical processes evolve in complex systems.
Recently, some models have been proposed to explain specific
aspects of heterogeneous dynamics in networks~\cite{Stehle2010,Jo2011,Karsai2012,Perra2012,Bagrow2013,Barrat2013,Starnini2013c,Starnini2014,Karsai2014},
most notably the empirically observed burstiness  (apparent in the broad distributions of inter-contact durations), known to
have a strong impact on dynamical processes on temporal networks~\cite{Vazquez2007,Miritello2011a,Karsai2011b}.
Memory effects have been proposed as an explanation of the emergence of heterogeneous dynamics~\cite{Stehle2010,Karsai2012,Karsai2014}, yet memory in human interaction dynamics has only been observed directly in the specific case of how mobile phone users connect to new contacts~\cite{Karsai2014}.
Furthermore, the individual role of different elementary memory mechanisms in the emergence of heterogeneous dynamics has not been investigated.

We here present a minimal yet general modeling framework consisting of a fixed set of agents (or nodes) that stochastically create and break contacts.
We investigate the rates of creation and deletion of contacts in empirical temporal networks, describing on the one hand face-to-face contacts and on the
other hand email communications, and find that long-term memory effects are present. We thus
propose four distinct memory mechanisms corresponding to these effects and
study systematically their individual and combined effects on the network model's dynamics and on dynamical processes taking place on the network.
We focus in particular on the distributions of the durations of contacts between two agents $i$ and $j$,
$p(\tau_{(i,j)})$, of the inter-contact durations for a given agent $i$ or a given link $(i,j)$,
$p(\Delta\tau_{i})$ and $p(\Delta\tau_{(i,j)})$ respectively, and finally of the numbers of contacts per link, $p(n)$,
which also has a crucial impact on the outcome of spreading processes on temporal networks~\cite{Gauvin2013,Karsai2014}.

The remainder of this paper is organised as follows.
Section~\ref{sec:results} presents the main results of the paper; namely, it first
describes the general modeling framework, then
shows the presence of long-term memory effects in empirical data, and
finally investigates how these affect network dynamics and dynamical processes on the networks. We discuss the results
and present some conclusions in
Section~\ref{sec:discussion}, while technical details are given in two appendices.
Appendix~\ref{sec:MasterEq} shows how analytical results for the dynamics of model networks are derived.
In Appendix~\ref{sec:epidemics}, we present details on the simulations of epidemic processes
and describe the quantities used for the quantitative comparison of the outcome of these simulations
on model and empirical temporal networks.
Appendix~\ref{sec:supfigs} contains supplementary figures.

\section{Results}
\label{sec:results}
%\paragraph{The model.}
\subsection{The model}
\label{sec:Rules}
We consider a population of $N$ agents. The $N(N-1)/2$ pairs $(i,j)$ of agents are all potential {\em links}.
If $i$ and $j$ are in contact the link $(i,j)$ is {\em active}, while $(i,j)$ is {\em inactive} when $i$ and $j$ are not in contact. The number of active links
at time $t$ is denoted $M_1(t)$.
Agents are characterized by the time $t-t_i$ elapsed since the last time $t_i$ they changed state, i.e., the last time that the agent either gained or lost a contact.
Links are characterized by their age, defined as the time $t-t_{(i,j)}$ elapsed since the link was either activated or inactivated.
%Agents and links are characterized by their {\bf ``age''}, where an agent's {\bf age} is defined as the time $t-t_i$ elapsed since the last time $t_i$ at which the agent $i$ either gained or lost a contact, and a link's {\bf age} is the time elapsed since the link was last either activated or inactivated, $t-t_{(i,j)}^{(1)}$ and $t-t_{(i,j)}^{(0)}$, respectively.
We initialize the network with all agents isolated (all links inactive).
At each time step $dt$, all active links and all agents are updated as follows:

(i) Each active link $(i,j)$ is inactivated with probability $dt\,z\,\fl(t-t_{(i,j)})$, where $\fl$ may depend on $t-t_{(i,j)}$ and $z$ is a parameter of the model controlling the rate with which contacts end;

(ii) Each agent $i$ initiates a contact with another agent with
probability $dt\,b\,\fa(t-t_i)$, where $\fa$ may depend on $t-t_i$ and $b$ is a parameter that controls the rate of contact creation.
The other agent $j$ is chosen among agents that are not in contact with $i$,
with probability $\Pa(t-t_j)\Pl(t-t_{(i,j)})$, where $\Pa$ may depend on $t-t_j$ and $\Pl$ on $t-t_{(i,j)}$.
If a link $(i,j)$ has never been active we set $t_{(i,j)}=0$.

%\paragraph{Memory.}
\subsection{Relating the model's ingredients to memory effects in empirical data}

In the model,
the ``memory kernels" $\fl$ and $\Pl$ control the rates with which a link is inactivated and activated, respectively, and $\fa$ and $\Pa$ control the rate with which agents enter new contacts, i.e., create or receive a contact, respectively. Specifically,
we show in Appendix \ref{sec:MasterEq} that
\begin{itemize}
\item the rate $r_{-,(i,j)}(t-t_{(i,j)})$ at which contacts of age $t-t_{(i,j)}$
    end is proportional to $\fl(t-t_{(i,j)})$,
\item the rate $r_{+,(i,j)}(t-t_{(i,j)})$ at which inactive links of age $t-t_{(i,j)}$ are activated is approximately proportional to $\Pl(t-t_{(i,j)})$,
\item the rate $r_{+,i}(t-t_i)$ at which agents that have not changed state since $t_i$ enter a new contact at $t$ is proportional to a linear combination of $\fa(t-t_i)$ and $\Pa(t-t_i)$
\end{itemize}
[see Eqs.~(\ref{eq:r_-,(i,j)})--(\ref{eq:r_+,i}) of Appendix~\ref{sec:MasterEq} for exact relations].

\begin{figure}[tb]
  \includegraphics[width=0.49\textwidth]{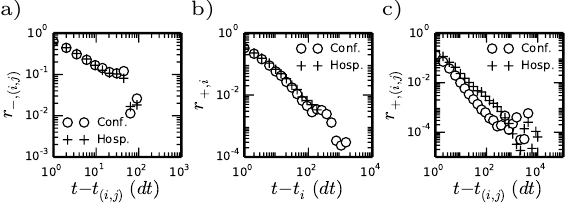}
  \caption{Memory effects in empirical data of face-to-face contacts recorded at a conference (conf.)~\cite{Stehle2011}
  and in a hospital ward (hosp.)~\cite{Vanhems2013a} (see Fig.~\ref{supfig:memory} for the case of an email exchange network)~\cite{SM}.
  a) Average rate $r_{-,(i,j)}$ at which contacts of {age} $t-t_{(i,j)}$ end, calculated as the number of links being inactivated at {age} $t-t_{(i,j)}$ divided by the total number of active links of {age} $t-t_{(i,j)}$. %; the rate $\Delta m_1(\tau)/m_1(\tau)$ estimates $z\,\fl(\tau)$.
 % corresponding to $z\,\fl$ in the model.
  b) Average rate $r_{+,i}$ at which agents that have been isolated for exactly $t-t_i$ enter a new contact, calculated as the number of agents that enter a contact after having been isolated for a time $t-t_i$ divided by the total number of agents that have been isolated for $t-t_i$.
  %; $\Delta n_0(\tau)/n_0(\tau)$ estimates $b\,\fa(\tau)+r_+\Pa(\tau)$.
  %corresponding to $b\,\fa+c\,r_+\Pa$ in the model.
  c) Average rate $r_{+,(i,j)}$ at which inactive links of {age} $t-t_{(i,j)}$ are activated, calculated as the number of links being activated at {age} $t-t_{(i,j)}$ divided by the total number of inactive links of {age} $t-t_{(i,j)}$. %; $\Delta m_0(\tau)/m_0(\tau)$ estimates
%  corresponding to $d\, \lambda_+\Pl$ in the model.
  %Here $r_+$ is the average rate with which an agent initiates a contact, $\lambda_+$ describes correlations between agents and links (see text), and $c$ and $d$ are constants which normalize $\Pa$ and $\Pl$.
%  Here $r_+(\tau)$ is the average rate with which a node of {\bf age} $\tau$ or lower initiates a contact and $r_+\equiv r_+(\infty)$.
  All quantities were averaged over all agents/links in the system and over the full recording time (approx. 32 hours for the conf. and 5 days for the hosp. data, recorded in $dt=20\ \mathrm{s}$ intervals).}
  \label{fig:memory}
\end{figure}

The simplest case where these rates are constant
as functions of $t-t_i$ and $t-t_{(i,j)}$  is thus obtained for $\fl=\fa=\Pa=\Pl=1$. It leads to a memoryless network
(``0" in the figures) with Poissonian statistics~(see also the numerical simulations in Section~\ref{subsec:IID}).
Figure \ref{fig:memory}
however shows
that the rates at which links are activated and inactivated in empirical temporal networks
 actually depend on $t-t_i$ and $t-t_{(i,j)}$,
with slowly decaying forms  (close to power laws with exponents close to one) providing evidence of long-term memory effects.

\subsection{Incorporating the memory effects into the model}
In order to capture the memory effects found in empirical data, we propose four memory mechanisms, each controlled by one of the model's memory kernels:
{\em contact self-reinforcement} (CSR) [$\fl(\tau)\sim\tau^{-\gamma}$]---the longer a contact has lasted, the less likely it is to end \cite{Stehle2010};
{\em activity self-reinforcement} (ASR) [$\fa(\tau)\sim\tau^{-\gamma}$]---the more recently an agent has been active, the more likely it is to initiate a new contact;
{\em agent-centric preferential attachment} (APA) [$\Pa(\tau)\sim\tau^{-\gamma}$]---the more recently an agent has been active, the more likely it is to be chosen by another agent initiating a contact;
{\em link-centric preferential attachment} (LPA) [$\Pl(\tau)\sim\tau^{-\gamma}$]---an agent initiating a contact is more likely to choose an agent
 it has recently been in contact with.
Note that, inspired by Fig.~\ref{fig:memory},
we will mostly consider slowly decaying functional forms for the memory kernels, and in particular $\sim \tau^{-1}$ in the numerical investigations.
We however emphasize that the framework and analytical computations remain valid for different functional forms (Appendix~\ref{sec:MasterEq} and Fig.~\ref{supfig:exponents})~\cite{SM}.
Slowly decaying kernels correspond to {\em rich-get-richer} mechanisms, i.e., self-reinforcing effects, akin to the preferential attachment
mechanism~\cite{Barabasi1999,Stehle2010}.
ASR and APA capture the idea that highly active agents tend to create more new contacts and are more attractive to other agents initiating new contacts;
CSR and LPA can be seen as crude models of ``friendship'', where CSR captures the reinforcement of contacts with time and LPA captures that one tends to interact more often with close acquaintances.
While Fig. \ref{fig:memory} shows that several memory effects are combined in empirical data,
our modeling framework allows us to explore their individual roles in the next subsection.

\begin{figure*}[!]
  \includegraphics[width=\textwidth]{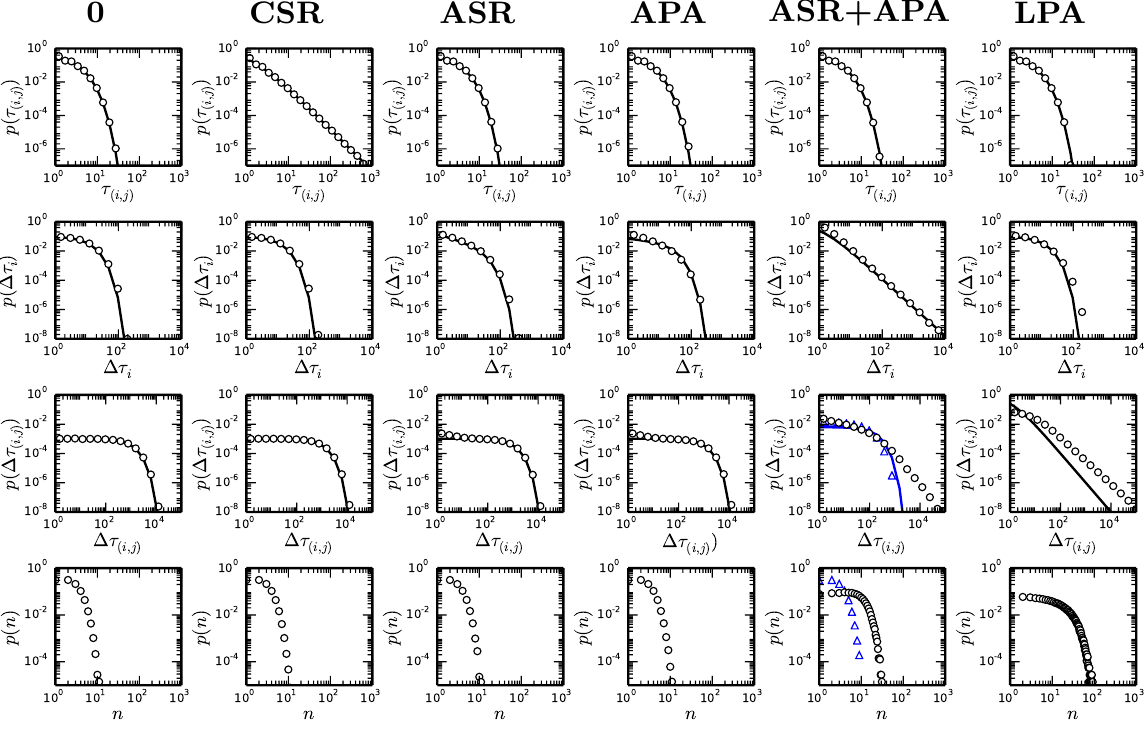}
  \caption{(Color online) Effect of individual memory mechanisms on dynamics: distributions of contact durations $\tij$, agents' inter-contact durations $\dti$, links' inter-contact durations $\dtij$, and number $n$ of contacts per link in networks integrated over time-windows of $\Delta T=10$,000.
  Open circles and full lines mark numerical and analytical results, respectively.
  Titles indicate the memory effects included in each column:
  for the memoryless network (0), $\fl=\fa=\Pa=\Pl=1$;
  for { CSR}, $\fl(\tau)=(1+\tau)^{-1}$;
  for ASR, $\fa(\tau)=(1+\tau)^{-1}$;
  for APA, $\Pa(\tau)=(1+\tau)^{-1}$;
  for { ASR+APA}, $\fa(\tau)=(1+\tau)^{-1}$ and $\Pa(\tau)=(1+\tau)^{-1}$;
  for { LPA}, $\Pl(\tau)=(1+\tau)^{-1}$.
  For all simulations $N=100$, $dt=0.1$, and the model was run until a stationary state was reached before
  recording statistics; $z$ and $b$ were chosen such that the average rate for an agent to initiate a
  new contact was $r^*_+=0.05$ and the average rate for an existing contact to end was $r^*_-=0.5$ in the
  quasi-stationary state (as witnessed by the mean of $M_1(t)$ remaining constant over time), hence the average number of active links was $M_1^*=10$.
  For ASR+APA, $p(\dtij)$ and $p(n)$ are also shown for  $r^*_+=0.4$ (dense network with $M_1^*=80$,
  blue line and triangles) since the distributions are then qualitatively different; for $r_+^*=0.05$, $p(\dtij)$ here shows a fat tail, whose form, however, is not accessible analytically.
  LPA induces an approximately scale-free distribution of $\dtij$ as predicted theoretically; theory overestimates the exponent, possibly
  because $m_0(t,t-\dtij)$ does not have a stationary state.}
  \label{fig:elements}
\end{figure*}

\subsection{Relation between memory mechanisms and heterogeneous dynamics}
\label{subsec:IID}
%\paragraph{CSR induces heterogeneous contact durations.}
\subsubsection{CSR induces heterogeneous contact durations}
The distribution of contact durations $\tau_{(i,j)}$ is the number of contacts of {age} $\tau_{(i,j)}=t-t_{(i,j)}$ ending at a given instant $t$, i.e., $p(\tau_{(i,j)})\propto-\partial_{t} m_1(t,t-\tau_{(i,j)})$, where $\partial_t$ is the differential
operator $d/dt$ and $m_1(t,t-\tau_{(i,j)})$ is the number of active links of {age} $\tau_{(i,j)}$ at time $t$.
Heterogeneous mean-field theory (HMFT) gives $m_1(t,t-\tau_{(i,j)}) \propto \exp[-z\int_0^{\tau_{(i,j)}}\fl(u)du]$~(Appendix~\ref{sec:MasterEq}), and requiring $p(\tau_{(i,j)})$ to be normalized yields
\beq
  p(\tau_{(i,j)}) = z\,\fl(\tau_{(i,j)})e^{-z\int_0^{\tau_{(i,j)}} \fl(u)du} \es.
  \label{eq:p(tau_ij)}
\eeq
$\fl$ thus governs the distribution of contact durations $\tau_{(i,j)}$ and CSR [$\fl(\tau)\sim\tau^{-\gamma}$]
induces a broad distribution of $\tau_{(i,j)}$, while it does not affect other distributions of interest (Fig.~\ref{fig:elements}).
For
$\fl(\tau)=(1+\tau)^{-1}$ we recover a scale-free distribution of contact durations:
$p(\tau_{(i,j)})=z(1+\tau_{(i,j)})^{-(z+1)}$~\cite{Note1}.

%\paragraph{ASR+APA induces broad distributions of agents' inter-contact durations.}
\subsubsection{ASR+APA induces broad distributions of agents' inter-contact durations}
The agents' inter-contact durations $\Delta\tau_i$ are defined, for each agent $i$,
as the time elapsed between the end of a contact and the beginning of another one. Since temporal
networks of practical interest are sparse, we can in practice
approximate $\Delta\tau_i$ by the durations agents stay isolated.
The distribution of $\Delta\tau_i$ is then found as $p(\Delta\tau_i) \propto -\partial_{t} n_0(t,t-\Delta\tau_i)$, where $n_0(t,t-\Delta\tau_i)$ is the number of agents which have been isolated since $t-\Delta\tau_i$ at time $t$. Analytical computations (Appendix~\ref{sec:MasterEq}) and numerical simulations show that ASR or APA individually
 [i.e., $(\fa(\tau)\sim\tau^{-\gamma}, \Pa(\tau)=1)$ or
$(\fa(\tau)=1, \Pa(\tau)\sim\tau^{-\gamma})$] lead to exponentially distributed $\Delta\tau_i$, while
their combination $(\fa(\tau)\sim\tau^{-\gamma}, \Pa(\tau)\sim\tau^{-\gamma})$
 induce a broad distribution of $\Delta\tau_i$ (Fig.~\ref{fig:elements}). For instance,
assuming that $\Pa$ has the same functional form as $\fa$ leads to~(Appendix~\ref{sec:MasterEq})
\beq
  p(\Delta\tau_i) = 2b\,\fa(\Delta\tau_i) e^{-2b\int_0^{\Delta\tau_i} \fa(u)du} \es.
  \label{eq:p(dtau_i)}
\eeq
For $\Pa(\tau)= \fa(\tau)=(1+\tau)^{-1}$, we obtain $p(\Delta\tau_i)=2b(1+\Delta\tau_i)^{-(2b+1)}$.

%\paragraph{LPA induces heterogeneous distribution of links' inter-contact durations.}
\subsubsection{LPA induces heterogeneous distribution of links' inter-contact durations}
\begin{figure*}[htb]
  \includegraphics[width=\textwidth]{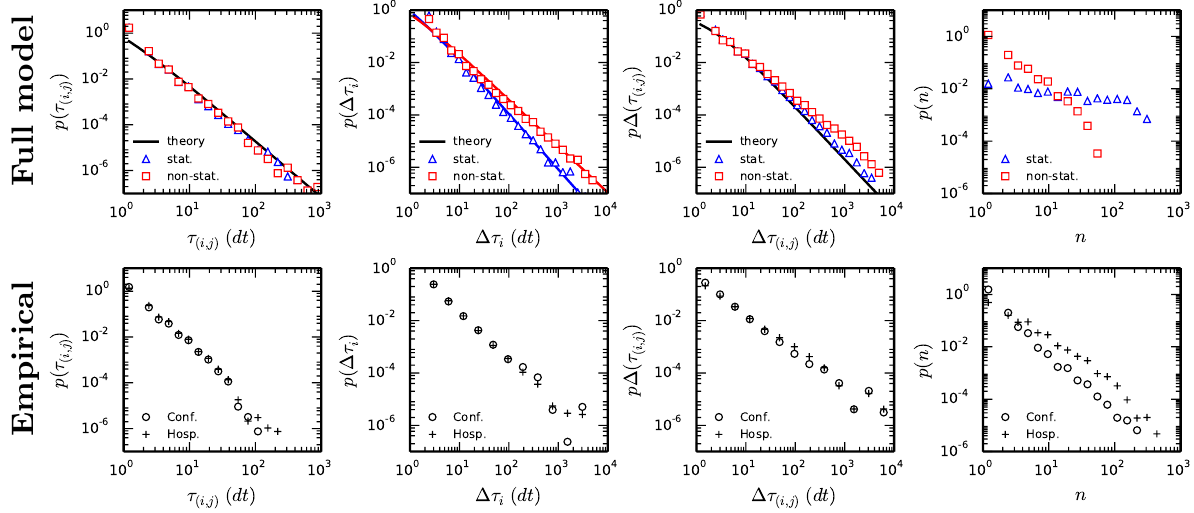}
  \caption{(Color online) Comparison of the properties of a synthetic network produced by our modelling framework including
  all four memory mechanisms (Full model) and of empirical (Empirical) face-to-face contact networks.
  The number of agents in the conference (conf.) data was $N=399$, and $N=80$ for the hospital (hosp.) data (see Table~\ref{suptab:stats} for summary statistics of empirical and model networks).
  Twenty realizations of the model networks were simulated for $T=5750$ time-steps $dt$, with $dt=1$, corresponding to $32$
  hours with $dt=20\ {\rm s}$ (the same as for the conf. data); for stationary networks (stat.) the model was run until the system
  reached a stationary state before statistics   were recorded, while contacts were recorded right
  away for non-stationary networks (non-stat.) (see Fig.~\ref{supfig:activity} for examples of the longitudinal activity of model and empirical networks)~\cite{SM}.
  $N=399$ was chosen to match the conf. data, and $b$ and $z$ were chosen such that the total number of contacts and their cumulated duration
  were approximately the same as in the conf. data ($b=0.56$ and $z=1.44$ for stationary networks, and $b=0.385$ and $z=1.435$ for non-stationary networks).
  See Fig.~\ref{supfig:KL_dist} for a quantitative comparison of the distributions corresponding to model and empirical networks~\cite{SM}.}
  \label{fig:model}
\end{figure*}

The distribution of links' inter-contact durations $\Delta\tau_{(i,j)}$, i.e., of the time elapsed between consecutive contacts of two given agents $i$ and $j$, is found as the rate of change $p(\Delta\tau_{(i,j)})\propto-\partial_t m_0(t,t-\Delta\tau_{(i,j)})$ of the number of inactive links of {age} $\Delta\tau_{(i,j)}$. This gives~(Appendix~\ref{sec:MasterEq})
\beqa
  p(\Delta\tau_{(i,j)},t) &=& 2d(t)\lambda_+(\Delta\tau_{(i,j)},t)\,\Pl(\Delta\tau_{(i,j)})/N \label{eq:p(dtau_ij)}\\
  &&\times e^{-\int_0^{\Delta\tau_{(i,j)}} \frac{2d(t-\tau+u)\lambda_+(u,t-\tau+u)\,\Pl(u)}{N}\,du} \es, \nonumber
\eeqa
where $d(t)$ normalizes $\Pl$ and $\lambda_+(\Delta\tau_{(i,j)},t)$ captures the effect of correlations between the time elapsed since an agent last changed state and the ages of its inactive links~(Appendix~\ref{sec:MasterEq}).
From Eq.~(\ref{eq:p(dtau_ij)}) we see that LPA [$\Pl(\tau)\sim\tau^{-\gamma}$] induces a
heterogeneous distribution of $\Delta\tau_{(i,j)}$ \cite{Note2}.
For $\Pl(\tau)=(1+\tau)^{-1}$ we obtain the simple scale-free distribution $p(\Delta\tau_{(i,j)})=\alpha(1+\tau)^{-(\alpha+1)}$;
numerical simulations show that $\alpha\approx0.5$ for a network with only LPA  (Fig.~\ref{fig:elements}) and $0.5\lessapprox\alpha\lessapprox1$ for a network with all four memory mechanisms (Figs.~\ref{fig:model}, \ref{supfig:exponents}, and \ref{supfig:hosp})~\cite{SM}, while HMFT gives $\alpha\approx1$ for a stationary network~(Appendix~\ref{sec:MasterEq}).

\begin{table}
  \centering
  \caption{Summary of the effects of the various
  memory mechanisms. Each column corresponds to a
  given mechanism and marks whether the resulting distribution (rows)
  is homogeneous/narrow (hom.) or heterogeneous/broad (het.).
  }
  \label{tab:summary}
  \begin{tabular}{ccccc}
  \hline
  \hline
                            & {\bf CSR}   & {\bf ASR+APA} & {\bf LPA} & {\bf All}\\
  \hline
    $p(\tau_{(i,j)})$               & het. & hom. & hom. & het.\\
    $p(\Delta\tau_{i})$   & hom. & het. & hom. & het.\\
    $p(\Delta\tau_{(i,j)})$   & hom. & hom./het.\tablenote{For a dense network $p(\Delta\tau_{(i,j)})$ is homogeneous; for a sparse network $p(\Delta\tau_{(i,j)})$ is heterogeneous.}
    & het. & het.\\
    $p(n)$                  & hom. & hom. & hom. & hom./het.\tablenote{$p(n)$ is homogeneous for a stationary network and heterogeneous for a non-stationary network.}\\
%  \hline
%  {   Region of}     & $z>1$     & $b>1/2$    &   all  & $z>1$, \\
%  {   stationarity\tablenote{Parameter values for which the expected value of $M_1(t)$ converges to a finite value.} }     &&&&   $b>1/2$         \\
  \hline
  \hline
  \end{tabular}
\end{table}

%\paragraph{CSR, ASR, APA, and LPA collectively induce heterogeneous numbers of contacts per link.}
\subsubsection{CSR, ASR, APA, and LPA collectively induce heterogeneous numbers of contacts per link}
None of the four mechanisms individually induce a broad $p(n)$
(Fig.~\ref{fig:elements}); however, including all of them in the model yields
a broad distribution of $n$ (Figs.~\ref{fig:model}, \ref{supfig:exponents}, and \ref{supfig:hosp})~\cite{SM}.

Table~\ref{tab:summary} summarizes the effects of the various memory mechanisms on network dynamics.

Besides broad distributions of $p(\tau_{(i,j)})$, $p(\Delta\tau_i)$, $p(\Delta\tau_{(i,j)})$, and $p(n)$,
the full model networks also present heterogeneous distributions of the lifetime of triangles
and agent activity levels (Fig.~\ref{supfig:stats})~\cite{SM}; moreover, when the networks are aggregated over time they present
heterogeneous distributions of edge weights, homogeneous degree distributions, and high clustering coefficients,
similarly to aggregated empirical temporal networks (Fig.~\ref{supfig:stats} and Table~\ref{suptab:stats}, see also Figs.~\ref{supfig:elements} and \ref{supfig:asr+apa} and  for other model networks)~\cite{SM}.

\begin{figure}[htb]
  \centering
  \includegraphics[width=0.5\textwidth]{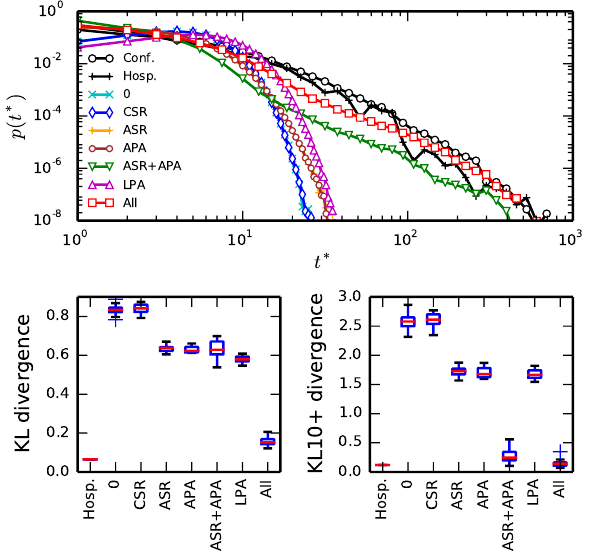}
  \caption{(Color online) Top: Log-binned distributions $p(t^*)$ of arrival times $t^*$ of the SI process at the nodes of empirical networks and of model networks.
  Bottom: Boxplots of Kullback-Leibler (KL) divergences between $p(t^*)$ on
  model networks and on the conf. network, and of KL10+ divergences
  where only values $t^*>10$ were taken into account;
  the KL and KL10+ divergence between the hosp. and conf. data are also shown for comparison.
  %(0) refers to a memoryless network.
  Distributions are the result of SI processes initiated at $200$ different times for all nodes ($1000$ times for empirical networks).
  For the model networks, we ran the SI process on $20$ different realizations of the network.
  All model networks consisted of $N=399$ nodes and were simulated for $T=5750\,dt$.
  The parameters $b$ and $z$ were chosen to approximately match the total number and duration of all contacts of the conf. network.
  Boxes mark quartiles (Q1 and Q3 in blue, median in red);
  whiskers extend beyond boxes to the most extreme points within 1.5 times the interquartile range; plus signs mark outliers.
  Networks with ASR+APA and all four memory mechanisms (All) are non-stationary, while other model networks are stationary.}
  \label{fig:SI}
\end{figure}

%\paragraph{Dynamical processes on temporal networks.}
\subsection{Dynamical processes on temporal networks}
To investigate the effect of the memory mechanisms, and resulting network heterogeneities, on dynamic processes taking place on
temporal networks, we simulate the paradigmatic deterministic Susceptible-Infected (SI) process
on the temporal networks obtained through our modeling framework. SI processes have indeed
often been used as simple but useful probes of temporal networks.
For each run, we start with all agents in the susceptible (S) state,
select at random a single agent as ``seed", and put it in the infectious (I) state. The ``infection" spreads then deterministically from infectious to susceptible agents
at each contact %: in other words, at each contact between a susceptible and an infectious agents, the susceptible becomes infectious
(see Appendix~\ref{sec:epidemics} for more details).
We repeat the process for all possible seeds and for different seeding times and we
consider for each agent the arrival time of the infection, i.e., the time at which the agent becomes infectious, and the distributions of these
times. The arrival times are measured
using {\em activity clocks}~\cite{Gauvin2013,Panisson2012}: the arrival {\em activity time} $t^*$ is defined for each agent
as the number of contacts it has taken part in from the start of the process to the arrival of the infection at the corresponding node.
This definition leads to
similar distributions of arrival times for different seeding times of the SI process and for seeds with different activity levels, and provides
an interesting tool for the comparison of models~\cite{Panisson2012,Gauvin2013}.
We provide a quantitative comparison of SI processes unfolding on model temporal networks and on empirical data sets by
computing the symmetrized Kullback-Leibler (KL) divergence between the corresponding arrival time distributions $p(t^*)$~(Appendix~\ref{sec:epidemics}),
an information-theoretic measure of the distance between two distributions related to the information lost when approximating one by the other~\cite{Kullback1951}.
We also consider KL divergences $\KL10+$ restricted to the tails of the distributions ($t^*>10$).

Figure~\ref{fig:SI} shows the arrival time distributions $p(t^*)$,
as well as boxplots of the KL divergences between $p(t^*)$  obtained from different realizations
of the network model and from the conf. temporal network (see Fig.~\ref{supfig:SI} for comparison with the hosp. data)~\cite{SM}.
Taken individually, each memory mechanism has little effect on $p(t^*)$;
CSR even has no effect at all since it only changes $p(\tau_{(i,j)})$, which has no effect on a deterministic SI process~\cite{Gauvin2013}.
More interestingly, on a network with ASR and APA, the tail of $p(t^*)$ is highly similar to the one obtained for spreading on
empirical networks; this suggests a crucial role of the burstiness of agents' contacts [exposed by
$p(\Delta\tau_{i})$], which is controlled by ASR+APA, in determining the occurrence of large $t^*$ values and
overall for diffusion phenomena on temporal networks~\cite{Vazquez2007,Karsai2011b,Starnini2012}.

The addition of LPA (``All" in Fig.~\ref{fig:SI}) is necessary to obtain a distribution of
arrival times $t^*$  that resembles that of the empirical data.  This confirms the importance of $p(n)$ in determining the global shape of $p(t^*)$,
as shown in~\cite{Gauvin2013}, and in particular for the distribution of the small $t^*$ values:
the comparison of $p(t^*)$ obtained for non-stationary and stationary model networks (Fig.~\ref{supfig:SI})~\cite{SM} highlights its sensitivity
to the precise shape of $p(n)$~\cite{Gauvin2013}. On the other hand, link burstiness by itself, as
determined by a broad  $p(\Delta\tau_{(i,j)})$, does not influence the spreading dynamics much (LPA in Fig.~\ref{fig:SI}), at least for the
deterministic SI process considered here.

\section{Discussion}
\label{sec:discussion}
The modeling framework presented here enabled us to characterize the emergence of several aspects of the temporal heterogeneities
observed in complex networks as a result of specific memory mechanisms controlling the creation and ending of contact events.
The four memory mechanisms proposed collectively result in broad distributions of contact and inter-contact durations and
of the number of contacts per link. They also yield
a broad distribution of agents' activities and may thus explain the emergence of heterogeneous activity observed recently
in large scale communication networks~\cite{Karsai2012,Karsai2014,Liu2014}.
The framework exposed here
also let us study the individual impact of each type of heterogeneity on dynamical processes on temporal networks.
In particular, broad distributions of
$\Delta\tau_{i}$ and $\Delta\tau_{(i,j)}$ are often interchangeably referred to as burstiness.
However, while both distributions are broad in empirical data, they are not identical. % [$p(\Delta\tau_{(i,j)})$ is broader than $p(\Delta\tau_i)$].
Furthermore, our study shows that these heterogeneous distributions may have different origins and have a different impact on
the unfolding of dynamical processes on the network.
Finally, our modelling framework highlighted the crucial roles of $p(\Delta\tau_{i})$ and $p(n)$ in spreading processes,
in agreement with~\cite{Gauvin2013}.

Our modeling framework can easily be extended to include more features such as circadian rhythms or the existence of groups of agents, such as classes or departments in school~\cite{Stehle2011a} or workplace~\cite{Genois2014} settings.
Considering groups of agents with different properties
may also prove a convenient way to model activity variations, which result
from agents entering or leaving the system, and thus to study how such variations might induce incomplete sampling of a population
and possible biases in empirical data.
Finally, it would be interesting to investigate the impact of the memory mechanisms presented here on other dynamic processes on
networks and in particular on more realistic epidemic models.

\section*{Acknowledgements}
The authors thank L. Gauvin and A. Panisson of the ISI Foundation, Turin, Italy, for fruitful discussions pertaining to the
analysis of SI epidemics on temporal networks. The authors also thank the SocioPatterns collaboration for privileged access to the conference data set.
This work was supported by by the EU FET project MULTIPLEX 317532 (to C.L.V. and A.B.) and the French ANR HARMS-flu, ANR-12-MONU-0018, (to M.G. and A.B.).

%=====================================
% Appendix:
\appendix
%=====================================
\begin{table*}
  \centering
  \caption{Summary statistics for model and empirical networks intergrated over the full time of recordings $T$: number $N_{\rm c}$ of connected (not isolated) nodes, number $E$ of edges, average degree $\Ein{k}$, average clustering coefficient $C$, and average clustering coefficient $C_{\rm ER}$ of a random (Erd\"os-R\'enyi) network with $N$ nodes and $E$ edges).  For the conf. dataset (and corresponding model networks) $N=399$ and $T=5$,$750\ dt$ and for the hosp. dataset  $N=80$ and $T=17$,$383\ dt$.
  Note that clustering coefficients from different networks are not directly comparable since the number of edges may differ; instead one should compare the ratios $C/C_{\rm ER}$. ASR+APA increases clustering, while other mechanisms have no significant effect.
  ASR+APA and LPA both reinforce specific links at the expense of others and lead to a decrease in the number of edges in the integrated networks.}
  \label{suptab:stats}
  \begin{tabular}{cccccc|ccccc}
  \hline
  \hline
  & && Conf.  && &&& Hosp. &&\\
  & $N_{\rm c}$ & $E$ & $\Ein{k}$ & $C$ & $C_{\rm ER}$ & $N_{\rm c}$ & $E$ & $\Ein{k}$ & $C$ & $C_{\rm ER}$ \\
  \hline
  0 & $399\pm0$ & $28200\pm100$ & $141\pm1$ & $0.356\pm0.002$ & 0.355 & $80\pm0$ & $3148\pm3$ & $78.7\pm0.08$ & $0.996\pm0.001$ & $0.996$ \\
  CSR & $399\pm0$ & $28200\pm100$ & $141\pm1$ & $0.356\pm0.002$ & 0.356 &    $80\pm0$ & $3147\pm3$ & $78.7\pm0.08$ & $0.996\pm0.001$ & $0.996$  \\
  ASR+APA  & $399\pm0$ & 22$500\pm700$ & $113\pm4$ & $0.44\pm0.01$ & $0.284$ & $80.0\pm0.2$ & 2$400\pm100$ & $59\pm3$ & $0.84\pm0.02$ & $0.75$  \\
  %(stat.) & $170\pm10$ & 10$000\pm1$,$000$ & $51\pm6$ & $0.83\pm0.01$ & $0.127$ & $37\pm3$ & $560\pm100$ & $14\pm2$ & $0.92\pm0.02$ & $0.18$  \\
  LPA & $399\pm0$ & $17500\pm1300$ & $88\pm7$ & $0.22\pm0.01$ & 0.22 & $80\pm0$ & $1580\pm20$ & $39.4\pm0.4$ & $0.50\pm0.01$ & $0.50$ \\
  All & $398\pm1$ & 9$600\pm300$ & $48\pm1$ & $0.178\pm0.004$ & $0.121$ & $79.8\pm0.4$ & 1$160\pm60$ & $29\pm2$ & $0.48\pm0.02$ & $0.37$ \\
  %(stat.) & $73\pm6$ & $304\pm40$ & $1.73\pm0.03$ & $0.24\pm0.04$ & $0.002$ &  $16\pm3$ & $67\pm17$ & $1.7\pm0.4$ & $0.75\pm0.05$ & $0.010$ \\
  Empirical & $396$ & 12265 & 62 & 0.32 & 0.16 &  $80$ & 1405 & 35 & 0.69 & $0.44$ \\
  \hline
  \hline
  \end{tabular}
\end{table*}

\section{Heterogeneous mean-field theory in the continuous limit}
\label{sec:MasterEq}
Here we derive heterogeneous mean-field (HMF) master equations governing the time evolution
of the network of connections between agents.
We consider the number $n_k(t,t')$ of nodes (agents) of degree $k$ that last changed state at time $t'$
and the number $m_1(t,t')$ and $m_0(t,t')$ of active and inactive links of {age} $t-t'$ at $t$. %(Note that we for convenience use a slightly different notation here than in the main text, e.g., $m_1(t,t')$ instead of $m_1(t-t',t)$ for the number of links of {\bf age} $t-t'$ at $t$.)
We assume that $t$, $t'$, $m_1(t,t')$, $m_0(t,t')$, and $n_k(t-t')$ are large enough to approximate them by continuous variables.

Section~1 derives HMF master equations for $m_1(t,t')$, $m_0(t,t')$, and $n_k(t-t')$.
Section~2 shows how these master equations simplify for a stationary system.
Section~3 gives the relation between the rates of creation and deletion of contacts and the memory kernels in the model, allowing
us to relate empirical measures to the model's ingredients.
Section~4 derives analytical expressions for the distributions of contact and inter-contact durations.
Section~5 derives expressions for these distributions for discretely sampled data.
Section~6 finally considers a selection of relevant examples.

\subsection{Derivation of master equations}
\label{A1}
\subsubsection{Updating a single contact}
We first consider the effect of updating a single contact as described in (i) in the main text. In the mean-field approximation%~\footnote{Since the network is small-world, the dimensionality of the system is effectively high and the mean-field approximation holds well.}
, this changes $m_1(t,t')$, $n_k(t,t')$, and $m_0(t,t')$ by
\beqa
  d m_1(t,t') &=& -dt\,z\,\fl(t-t')m_1(t,t')/M_1(t) \es,\\
  d n_k(t,t') &=& - 2dt\,\lambda_{-}(t-t',t)\frac{k}{\Ein{k}(t)}n_k(t,t')/N \\
  &&+ 2dt\frac{(k+1)}{\Ein{k}(t)}q_{-,k+1}(t)\delta(t-t') \es, \nonumber \\
  d m_0(t,t') &=& dt\,r_-(t) \delta(t-t') \es.
\eeqa
Here $M_1(t)=\int_0^t m_1(t,u)du$ is the total number of active links.
$\Ein{k}(t)$ is the average node degree.
$r_-(t)$ is the rate at which links are inactivated with
\beq
  r_-(t) = z\int_0^{t} \fl(t-u)\frac{m_1(t,u)}{M_1(t)} du \es.
  \label{eq:r_-(t,t')}
\eeq
$q_{-,k}(t)= \int_0^t r_-(t,u)n_k(t,u)du/N$.
Finally, $\lambda_-(t-t',t)$ is a term that describes correlations between the time $t-t'$ elapsed since a node last changed state and the ages of its active links; it captures that a node's links are all at least as old as $t-t'$ and that the age of at least one of the node's links is equal to $t-t'$; $\lambda_-(t-t',t)$ is given by
\beq
  \lambda_{-}(t-t',t) = z \int_0^{t'} \fa(t-u)\rho_{1}(t-u,t,t')\frac{m_1(t,u)}{M_1(t)}du \es,
\eeq
where $\rho_{1}(t-u,t,t')$ is proportional to the probability that a given active link of {age} $t-u$ is connected to a node which last changed state at time $t'$.

\subsubsection{Updating a single node}
Secondly, we consider the effect of updating a single node as described in (ii) in the main text. This changes $m_1(t,t')$, $n_k(t,t')$, and $m_0(t,t')$ as
\beq
  d m_1(t,t') = dt\, r_+(t)\delta(t-t') \es,
\eeq
\beq
  d n_0(t,t') = -dt\left[b\,\fa(t-t') + r_+(t)c(t)\Pa(t-t')\right]\frac{n_0(t,t')}{N} \es,
\eeq
\beqa
  d n_k(t,t') &=& -dt\left[b\,\fa(t-t') + r_+(t)c(t)\Pa(t-t')\right]\frac{n_k(t,t')}{N} \nonumber\\
   &&+ dt\left[r_{+,k-1}(t)+r_+(t)q_{+,k-1}(t)\right]\delta(t-t') \es, \nonumber\\
\eeqa
and
\beq
  dm_0(t,t') = - dt\,\lambda_+(t-t',t)d(t)\Pl(t-t')\frac{m_0(t,t')}{M_0(t)} \es.
\eeq
Here $r_+(t)$ is the rate at which nodes create contacts, with
\beq
  r_+(t) = (b/N)\int_{0}^t \fa(t-u)n(t,u) du \es,
  \label{eq:r_+(t,t')}
\eeq
where $n(t,u)=\sum_{k=0}^\infty n_k(t,u)$ is the total number of nodes which last changed state at time $u$.
$dt\,r_{+,k}(t)$ is the probability to add a contact to a node of degree $k$, with
\beq
  r_{+,k}(t) = (b/N)\int_0^t \fa(t-u)n_k(t,u)du \es.
\eeq
%$N_k(t)=\int_0^tn_k(t,u)du$ is the total number of nodes of degree $k$ at time $t$;
$c(t)$ is a normalization constant for $\Pa$.
$q_{+,k}(t) = c(t)\int_0^t \Pa(t-u)n_k(t,u)du/N$ and satisfies the constraint $\sum_{k=0}^\infty q_{+,k}(t) = 1$ since we require $\Ein{c\,\Pa}=1$.
$d(t)$ normalizes $\Pl$.
$M_0(t)=N(N-1)/2-M_1(t)$ is the total number of inactive links.
Finally, $\lambda_+(t-t',t)$ describes correlations between the time $t-t'$ elapsed since a node last changed state and the ages of its inactive links; it is given by
\beqa
  \lambda_+(t-t',t) &=& \int_{t'}^t [b\fa(t-u)+r_+(t,0)\Pa(t-u)]\nonumber\\
  &&\qquad\times\, \rho_0(t-t',t,u)\frac{n(t,u)}{N}\, du \es,
  \label{eq:lambda_+}
\eeqa
where $\rho_0(t-t',t,u)$ is proportional to the probability that a given inactive link of {age} $t-t'$ is connected to a node which last changed state at time $u$.
%in the limit of a very sparse network, nodes have at most one contact at a time, which then has the same age as the node, and $\lambda_+(t-t',t)$ is proportional to $b\,\fa(t-t')+r_+(t,0)c(t)\Pa(t-t')$; conversely, for a highly dense network, each node has many links, and $\lambda_+(t-t',t)$ is approximately equal to $r_+(t,t')$.

\subsubsection{Continuous-time master equations}
Each time-step, we update all $N$ nodes and $M(t)$ active links. The differential master equations governing the evolution of $m_1(t,t')$, $n_k(t,t')$, and $m_0(t,t')$ are thus in the continuous limit:
\beq
  \partial_t m_1(t,t') = - z\,\fl(t-t')m_1(t,t') + Nr_+(t)\delta(t-t') \es,
  \label{eq:d_tm1(t,t')}
\eeq
\beqa
  \partial_t n_0(t,t')
  &=& - \left[b\,\fa(t-t') + r_+(t)c(t)\Pa(t-t')\right]n_0(t,t') \nonumber\\
  &&\quad+ \pi_{1,0}(t)\delta(t-t')
  \label{eq:d_tN_0(t,t')} \es,\\
  \partial_t n_k(t,t')
  &=& - \left[b\,\fa(t-t') + r_+(t)c(t)\Pa(t-t')\right]n_k(t,t') \nonumber\\
  && \quad- k\lambda_-(t-t',t)n_k(t,t') \label{eq:d_tN_k(t,t')}\\
  && \qquad + \left[\pi_{k-1,k}(t)+\pi_{k+1,k}(t)\right]\delta(t-t') \es, \nonumber
\eeqa
and
\beqa
  \partial_t m_0(t,t') &=& -\frac{2\lambda_+(t-t',t)d(t)\Pl(t-t')}{N-1-2M_1(t)/N}m_0(t,t')
  \nonumber\\
  && \quad + M_1(t)r_-(t)\delta(t-t') \es, \label{eq:dm0(t,t')}
\eeqa
where we have used that $\Ein{k}(t)=2M_1(t)/N$. %Here $\partial_t$ denotes the partial differential operator w.r.t. $t$;
Here $\pi_{k-1,k}(t)$ and $\pi_{k+1,k}(t)$ are the number of nodes gaining or losing a link during a time-step, respectively~\cite{Note3}, they are equal to
\beqa
  \pi_{k-1,k}(t) &=& N[r_{+,k-1}(t) + r_+(t)q_{+,k-1}(t)] \es,\\
  \pi_{k+1,k}(t) &=& N(k+1)q_{-,k+1}(t) \es.
\eeqa

Integrating Eq.~(\ref{eq:d_tm1(t,t')}) w.r.t. $t'$ gives
\beq
  \partial_t M_1(t) = Nr_+(t) - M_1(t)r_-(t).
  \label{eq:dM_1(t)}
\eeq
Thus, in a stationary state we find $Nr_+(t)=M_1(t)r_-(t)$ as one should expect.

\subsection{Master equations simplify for a stationary system}
Depending on the form of $\fl$, $\fa$, $\Pa$, and $\Pl$ and the values of $b$ and $z$ the system may enter a stationary state after an initial transitory behavior. In the stationary state the master equations governing $m_1(t,t')$, $n_k(t,t')$, and $m_0(t,t')$ simplify.
If we assume that system has reached a stationary state, $m_1$, $n_k$, and $m_0$ only depend on $\tau=t-t'$.

The average number of active links is then constant,
\beq
  M_1=Nr^*_+/r^*_-\es,
  \label{eq:M_1}
\eeq
with
\beq
  r^*_{+} = (b/N)\int_0^\infty \fa(u)n^*(u) du \es,
\eeq
and
\beq
  r^*_- = (z/M_1^*)\int_0^\infty \fl(u) m_1^*(u)du \es
  \label{eq:r_-1}
\eeq
where the star denotes stationarity.

The number $m_1^*(\tau)$ of active links of {age} $\tau$ is governed by
\beq
  \partial_\tau m_1^*(\tau) = -z\,\fl(\tau)m_1^*(\tau) + Nr_+^*\delta(\tau)\es.
  \label{eq:d_tm(tau)}
\eeq
Equations~(\ref{eq:d_tN_0(t,t')}) and (\ref{eq:d_tN_k(t,t')}) give us master equations governing the number of nodes of degree $k$ which have remained in the same state for a time $\tau$,
\beqa
  \partial_\tau n^*_k(\tau) &=& -\left[b\,\fa(\tau)+r^*_+c^*\Pa(\tau)+k\lambda^*_-(\tau)\right]n^*_k(\tau)\nonumber\\
  &&\quad +n^*_k(0)\delta(\tau) \es. \label{eq:d_tn_k(tau)}
\eeqa
Finally, Eq.~(\ref{eq:dm0(t,t')}) reduces to
\beq
  \partial_\tau m_0(\tau) = - \frac{2\lambda_+^*(\tau)d^*\Pl(\tau)}{N-1-2r_+^*/r_-^*} m_0^*(\tau) + Nr_+^*\delta(\tau) \es.
  \label{eq:dm_0(tau)}
\eeq

Equation~(\ref{eq:d_tm(tau)}) gives
\beq
  m_1^*(\tau)=Nr_+^*e^{-z\int_0^\tau \fl(u)du} \es,
\eeq
and thus
\beq
  M_1^*=Nr_+^*\int_0^\infty e^{-z\int_0^\tau \fl(u)du} d\tau \es.
\eeq
So from Eq.~(\ref{eq:M_1}) we then have
\beq
  r_-^* = \left(\int_0^\infty e^{-z\int_0^\tau \fl(u)du} d\tau\right)^{-1} \es.
  \label{eq:r_-^*}
\eeq
%and, finally, from Eq.~(\ref{eq:r_-(tau)1}), that %$r_-^*(\tau)=r_-^*\,m_1^*(\tau)/(Nr_+^*)$.
%$r_-^*(\tau)=r_-^*\,\exp[-z\int_0^\tau \fl(u)du]$.

\subsection{Relation between rates of contact creation and deletion and memory kernels}
The relation between the rates of contact creation and deletion and the memory kernels of the model can in the HMF approximation be derived in a manner analogous to the master equations governing $m_1$, $m_0$, and $n_k$. We here use the results of Section~\ref{A1} [Eqs.~(\ref{eq:d_tm1(t,t')})--(\ref{eq:dm0(t,t')})].

The average rate $r_{-,(i,j)}(t-t')$ at which a contact ends is by definition given by $-\partial_t m_1(t,t')/m_1(t,t')$ for $t>t'$. Equation~(\ref{eq:d_tm1(t,t')}) thus directly gives
\beq
  r_{-,(i,j)}(t-t') = z\,\fl(t-t') \es.
  \label{eq:r_-,(i,j)}
\eeq
Since data is always discrete, $r_{-,(i,j)}(t-t')$ is estimated from data as $-\overline{\Delta m_1(t,t')}/\overline{m_1(t,t')}$, where $-\Delta m_1(t,t')$ is the number of contacts of age $t-t'$ ending at time $t$ and $\overline{\ldots}$ denotes the average over $t$.

Similarly, $r_{+,(i,j)}(t-t',t) = -\partial m_0(t,t')/m_0(t,t')$ for $t>t'$, and thus [Eq.~(\ref{eq:dm0(t,t')})]
\beq
  r_{+,(i,j)}(t-t',t) = \alpha(t-t',t)\Pl(t-t') \es, \label{eq:r_+,(i,j)}
\eeq
where $\alpha=2d(t)\lambda_+(t-t',t)/[N-1-2M_1(t)/N]$. Averaging Eq.~(\ref{eq:r_+,(i,j)}) over $t$ yields the average rate at which inactive links are activated:
\beq
  {r_{+,(i,j)}}(t-t') = \overline{\alpha(t-t',t)}\Pl(t-t') \es. %\label{eq:r_+,(i,j)}
\eeq
For discrete data, $r_{+,(i,j)}(t-t')$ is estimated as $-\overline{\Delta m_0(t,t')}/\overline{m_0(t,t')}$, where $-\Delta m_0(t,t')$ is the number of inactive links of age $t-t'$ being activated at time $t$.

Finally, the rate $r_{+,i}(t-t')$ at which agents enter into contact is the same as the rate at which agents change degree from $k$ to $k+1$, thus it is equal to, e.g., $-\partial_t n_0(t,t')/n_0(t,t')$ [Eq.~(\ref{eq:d_tN_0(t,t')})] and is given by
\beq
  r_{+,i}(t-t',t) = b\,\fa(t-t') + c(t)r_+(t)\Pa(t-t') \es.
  \label{eq:r_+,i}
\eeq
Averaging $r_{+,i}$ over $t$ then yields a linear combination of $\fa$ and $\Pa$, which can be estimated from data as $-\Delta \overline{n_0(t,t')}/\overline{n_0(t,t')}$, where $-\Delta n_0(t,t')$ is the number of isolated agents gaining a contact at time $t$.

\subsection{Distributions of contact and inter-contact durations}
\subsubsection{Contact durations}
The distribution of contact durations $\tij$ is given by the number of contacts of {age} $\tij=t-t'$ ending at a given instant, i.e., $p(\tij)\propto-\partial_{t} m_1(t,t')$.
From Eq.~(\ref{eq:d_tm1(t,t')}) we have
\beq
  m_0(t,t') = Nr_+(t')e^{-z\int_{t'}^t \fl(u-t')du} \es,
\eeq
which gives Eq.~(\ref{eq:p(tau_ij)}) by requiring $p(\tij)$ to be normalized.

\subsubsection{Agents' inter-contact durations}
 We find in the same manner as above the distribution of times during which a node stays isolated $\dti'$.
Note that $\dti'$ is in general not the same as the node's inter-contact durations $\dti$. However, since temporal networks of practical interest are sparse [$\Ein{k}(t)\ll1$] we use in the following the approximation $\dti=\dti'=t-t'$.
From Eq.~(\ref{eq:d_tN_0(t,t')}) we have
\beq
  n_0(t,t') = N q_{-,1}(t') e^{-\int_{t'}^t [b\fa(u-t')+r_+(u)c(u)\Pa(u-t')]du} \es.
\eeq
From $p(\dti,t) \propto -\partial_{t} n_0(t,t')$ we then obtain that
\beqa
  p(\dti,t) &=& [b\,\fa(\dti)+r_+(t)c(t)\Pa(\dti)] \label{eq:p_ab-ac}\\
  &&\times e^{-\int_0^{\dti} [b\,\fa(u)+r_+(t-\dti+u)c(t-\dti+u)\Pa(u)]du} \es.
  \nonumber
\eeqa
Note that $p(\dti,t)$ in general depends not only on $\dti$, but also on $t$; for $\Pa(\tau)= \fa(\tau)$, the requirement that $c\,\Pa$ is normalized ($\Ein{c\,\Pa}=1$) means, however, that $c(t)=b/r_+(t)$, and we recover Eq.~(\ref{eq:p(dtau_i)}).
Conversely, for a stationary network, $p(\dti)$ is by definition independent of $t$, and we here have in general
\beqa
  p(\dti) &=& [b\,\fa(\dti)+r_+^*c^*\Pa(\dti)] \label{eq:p_ab-ac-stat}\\
  &&\quad\times e^{-\int_0^{\dti} [b\,\fa(u)+r_+^*c^*\Pa(u)]du} \es.
  \nonumber
\eeqa

\subsubsection{Links' inter-contact durations}
Equation~(\ref{eq:dm0(t,t')}) gives us
%\beqa
%  m_0(t,t') &=& M_1(t')r_-(t') \\
%  &&\times e^{-\int_{t'}^t 2\lambda_+(u,t')c(u)\Pl(u-t')du/N} \es, \nonumber
%\eeqa
\beq
  m_0(t,t') = M_1(t')r_-(t') e^{-\int_{t'}^t 2\lambda_+(u,t')d(u)\Pl(u-t')du/N} \es,
\eeq
where we have used that $M_0(t)=N[N-1-\Ein{k}(t)]/2\approx N^2/2$. We then obtain from $p(\dtij,t) \propto -\partial_{t} m_0(t,t')$ that the links' inter-contact durations $\dtij=t-t'$ are distributed as given by Eq.~(\ref{eq:p(dtau_ij)}).
For a stationary network, $\lambda_+$ and $\Pa$ do not depend on $t$, and the expression for $p(\dtij)$ simplifies,
\beqa
  p(\dtij) &=& 2d^*\lambda_+^*(\dtij)\Pl^*(\dtij)/N  \nonumber\\
  &&\quad\times e^{-(2d^*/N)\int_0^\tau \lambda_+^*(u)\Pl^*(u)du} \es. \label{eq:p_ab-ab}
\eeqa

\subsection{Distributions for discrete data}
Empirical and simulated data are by nature discrete, whereas our theory is continuous. To relate the two, we derive here a discrete version of the distributions of (inter-)contact durations.

An event (contact or inter-contact duration) which lasts for a time $(k-1)dt<\tau\leq k\,dt$ is recorded as lasting $\tau_k=k\,dt$. The probability of recording an event that lasts $\tau=\tau_k$ is then given by
\beq
  p(\tau_k)=S([k-1]dt,t)-S(k\,dt,t) \es,
\eeq
where $S$ is the survival distribution of $\tau$, given by normalized versions of $m_1$, $n_0$, and $m_0$ for $\tij$, $\dti$, and $\dtij$, respectively [normalized such that $S(0)=1$]; $S$ may or may not depend on $t$.
In general, $S$ is given by
\beq
  S(\tau,t) = e^{-\int_0^\tau \beta(u,t-\tau)du} \es,
\eeq
where $\beta(u)=z\,\fl(u)$ for $\tij$, $\beta(u,t-\tau)=b\,\fa(u)+c(t-\tau+u)r_+(t-\tau+u,0)\Pa(u,t-\tau+u)$ for $\dti$, and $\beta(u,t-\tau)=2d(t-\tau+u)\lambda_+(u,t-\tau+u)\Pl(u,t-\tau+u)$ for $\dtij$.

\subsection{Case studies}

\subsubsection{Memoryless network  (0)}
We first consider the simplest possible network obtainable using our modeling framework, that of a memoryless network.
We set $\fa=\fl=\Pa=\Pl=1$, then $r_+(t)=b$ [Eq.~(\ref{eq:r_+(t,t')})] and $r_-(t)=z$ [Eq.~(\ref{eq:r_-(t,t')})].
Since the network is memoryless it rapidly reaches a stationary state where the average number of active links is $M_1^*=Nb/z$ [Eq.~(\ref{eq:M_1})]. From Eq.~(\ref{eq:p(tau_ij)}) we obtain:
\beq
  p(\tij) = ze^{-z\tij} \es.
  \label{eq:p_ab(0)}
\eeq
Equation~(\ref{eq:p(dtau_i)}) gives:
\beq
  p(\dti) = 2be^{-2b\dti} \es.
  \label{eq:p_ab-ac(0)}
\eeq
By approximating $r^*_+(\tau)$ by $r^*_+=b$ we finally obtain from Eq.~(\ref{eq:p_ab-ab}) in the stationary state:
\beq
  p(\dtij) = \frac{2b \exp\left(-\frac{2b\dtij}{N-1-2b/z}\right)}{N-1-2b/z}  \es.
  \label{eq:p_ab-ab(0)}
\eeq

Thus, $\tij$, $\dti$, and $\dtij$ are all exponentially distributed for a memoryless network~[Fig.~\ref{fig:elements}(0)]. %A slightly more involved derivation based on Eq.~(\ref{eq:d_tm(tau)}) shows that the distribution of instantaneous degrees $k$ in the memoryless network is Poissonian with $\Ein{k}=2b/z$, as in a random Erd\~{o}s-R\'{e}nyi (ER) network.
Considering that contacts are created and broken at random (uncorrelated) and with constant rates, we find that the distribution of instantaneous degrees $k$ of the nodes in the network must be Poissonian, with mean degree given by the identity $\Ein{k}=2M_1^*/N=2b/z$.

\subsubsection{Contact self-reinforcement (CSR)}
To obtain a network with CSR we set $\fl(\tau)=(1+\tau)^{-1}$ and let $\fa$, $\Pa$, and $\Pl$ be constant. Then
$r_+(t)=b$ [Eq.~(\ref{eq:r_+(t,t')})] and $r_-^*=z-1$ [Eq.~(\ref{eq:r_-^*})] in the stationary state; the number of active links is here $M_1^*=Nb/(z-1)$ for $z>1$ [Eq.~(\ref{eq:M_1})], while $M_1(t)$ diverges for $z\leq1$.
From Eqs.~(\ref{eq:p(tau_ij)})--(\ref{eq:p(dtau_ij)}) we see that CSR results in a scale-free distribution of $\tij$, here with
\beq
  p(\tij) = z(1+\tau)^{-(z+1)} \enspace,
  \label{eq:p(tij)}
\eeq
while  $\dti$ and $\dtij$ remain exponentially distributed as for the memoryless network [Fig.~\ref{fig:elements}(CSR)];
$p(\dti)$ is for a stationary network given by Eq.~(\ref{eq:p_ab-ac(0)}) and
\beq
  p(\dtij) = \frac{2b \exp\left(-\frac{2b\dtij}{N-1-2b/(z-1)}\right)}{N-1-2b/(z-1)} \es.
\eeq
The nodes' instantaneous degrees are Poisson distributed with $\Ein{k}=2b/(z-1)$.

\subsubsection{Activity self-reinforcement (ASR)}
To obtain a system with ASR we set $\fa(\tau)=(1+\tau)^{-1}$ and let $\fl$, $\Pa$, and $\Pl$ be constant. Here $r_-(t)=z$ [Eq.~(\ref{eq:r_-(t,t')})]. The system rapidly reaches a stationary state, where $M^*_1=Nr_+^*/z$ [Eq.~(\ref{eq:M_1})]. The distributions $p(\tij)$ and $ p(\dtij)$ are exponential [Fig.~\ref{fig:elements}(ASR)], with $p(\tij)$ given by Eq.~(\ref{eq:p_ab(0)}) and
\beq
  p(\dtij) = \frac{2r_+^* \exp\left(-\frac{2r_+^*\dtij}{N-1-2r_+^*/z}\right)}{N-1-2r_+^*/z} \es.
\eeq
Finally, Eq.~(\ref{eq:p_ab-ac-stat}) gives us:
\beqa
  p(\dti) &=& [b(1+\dti)^{-1}+r_+^*] (1+\dti)^{-b} \nonumber\\
  &&\quad\times e^{-r_+^*\dti} \es,
\eeqa
i.e., ASR has a small effect on $p(\dti)$ for small $\dti$, the exponential term (due to $\Pa=1$), however, dominates at large $\dti$ [Fig.~\ref{fig:elements}(ASR)].
ASR induces correlations in the creation of links, which means that we do not obtain a closed-form expression for the instantaneous degree distribution.

\subsubsection{Agent-centric preferential attachment (APA)}
To induce APA we let $\Pa(\tau)=(1+\tau)^{-1}$ and let $\fl$, $\fa$, and $\Pl$ be constant. Here $r_+(t)=b$ [Eq.~(\ref{eq:r_+(t,t')}] and  $r_-(t)=z$ [Eq.~(\ref{eq:r_-(t,t')}].
In this case the system also rapidly reaches a stationary state, where  $M^*_1=Nb/z$ [Eq.~(\ref{eq:M_1})]. The distributions $p(\tij)$ and $p(\dtij)$ are exponential as for the memoryless network [Fig.~\ref{fig:elements}(APA)] and are given by Eqs.~(\ref{eq:p_ab(0)}) and (\ref{eq:p_ab-ab(0)}), respectively.
Finally, we have from Eq.~(\ref{eq:p_ab-ac-stat}):
\beqa
  p(\dti) &=& b[c^*(1+\dti)^{-1}+1](1+\dti)^{-c^*b} \nonumber\\
  &&\quad\times e^{-b\dti} \es,
\eeqa
where numerical simulations show that $c^*\approx1$ [Fig.~\ref{fig:elements}(APA)].
As for ASR, we do not find a closed-form expression for the instantaneous degree distribution of a network with APA.

\subsubsection{ASR+APA}
By letting $f=(1+\tau)^{-1}$ and $\Pa(\tau)=(1+\tau)^{-1}$, and keeping $\fl$ and $\Pl$ constant, we obtain a network with ASR and APA.
Here $c(t)=b/r_+(t)$ and $r_-(t)=z$ [Eq.~(\ref{eq:r_-^*})]; the number of active links is in the stationary state $M_1^*=Nr_+^*/z$ for $b>1/2$ [Eq.~(\ref{eq:M_1})], while the system does not have a (non-empty) stationary state for $b\leq1/2$ as $M_1(t)$ tends to zero.
Since $\fl$ is constant, $p(\tij)$ is given by Eq.~(\ref{eq:p_ab(0)}) [Fig.~\ref{fig:elements}(ASR+APA)].
ASR+APA leads to a scale-free distribution of $\dti$ [Fig.~\ref{fig:elements}(ASR+APA)] given by
\beq
  p(\dti) = 2b(1+\dti)^{-(2b+1)} \es.
  \label{eq:p(dti)}
\eeq
Finally, ASR+APA can make $\lambda_+(t-t',t)$ a slowly decaying function of $\tau=t-t'$ through its dependency on $\fa$ and $\Pa$ [Eq.~(\ref{eq:lambda_+})]. Depending on the parameter values of the system, this may result in a heterogeneous distribution of $\dtij$ [Eq.~(\ref{eq:p(dtau_ij)}), and Figs.~\ref{fig:elements}(ASR+APA) and \ref{supfig:asr+apa}]~\cite{SM}; a simple analytical expression for the shape of $p(\dtij)$ is not available, however.

\subsubsection{Link-centric preferential attachment (LPA)}
LPA is induced by letting $\fl=\fa=\Pa=1$ and $\Pl(\tau)=(1+\tau)^{-1}$. Here $r_+(t)=b$, $r_-(t)=z$, and thus $M_1^*=Nb/z$ [Eq.~(\ref{eq:M_1})] in a stationary state.
The distributions $p(\tij)$ and $p(\dti)$ are given by Eqs.~(\ref{eq:p_ab(0)}) and (\ref{eq:p_ab-ab(0)}) [Fig.~\ref{fig:elements}(LPA)].
Assuming that $\lambda_+(t-t',t)$ and $d(t)$ change slowly with $t$ compared to $\dtij$ we can approximate $\lambda_+(u-t',t)$ by $\lambda_+(t-t',t)$ and $d(u)$ by $d(t)$ for $t'\leq u\leq t$; we then find that $p(\dtij,t)$ is given by
\beq
  p(\dtij,t) = \alpha(t,\dtij)(1+\dtij)^{-[\alpha(t,\dtij)+1]} \es,
  \label{eq:p(dtij,t,t')}
\eeq
with $\alpha(t,\dtij)=2d(t)\lambda_+(\dtij,t)/N$, where we have used that $N\gg 2M_1(t)/N+1$.
If we furthermore ignore node-link correlations, we can approximate $\lambda_+(t-t',t)$ by $r_+(t)$, and the above expression simplifies to
\beq
  p(\dtij,t) = \alpha(t)(1+\dtij)^{-(\alpha(t)+1)} \es.
  \label{eq:p(dtij,t)}
\eeq
If we finally assume that the system is stationary, we have [Eq.~(\ref{eq:dm_0(tau)})]
\beq
  m_0^*(\tau) = Nb(1+\tau)^{-\alpha^*} \es,
  \label{eq:m0(tau),LPA}
\eeq
where $\alpha^*\approx2b\,d^*/N$.
By integrating $m_0^*(\tau)$ we obtain
\beqa
  M_0^* &=& Nr_+^*\int_1^\infty u^{-\alpha^*}du \nonumber\\
  &=& Nr^*_+/(\alpha^*-1) \es,
\eeqa
for $\alpha^*>1$. Using that $M_0^*=N(N-1)/2-M_1^*$ we find $\alpha^*=1+Nr^*_+/M_0^*=1+2b/N\approx1$.
Numerical simulations show that $\overline{\alpha}\approx0.5$ [Fig.~\ref{fig:elements}(LPA)], where $\overline{\cdots}$ denotes an average over the simulations.

As for networks with ASR or APA, we do not obtain a closed-form expression for the degree distribution of a network with LPA.

\subsubsection{Full model}
We set $\fl(\tau)=\fa(\tau)=\Pa(\tau)=\Pl=(1+\tau)^{-1}$ to obtain a network with all four memory mechanisms (CSR, ASR, APA, and LPA).
%The normalization of $\Pa$ assures that $\Ein{\Pa(\tau)}=1$.
We here have that $r_-^*=z-1$ [Eq.~(\ref{eq:r_-^*})] and thus that $M^*_1=Nr_+^*/(z-1)$ in the stationary state [Eq.~(\ref{eq:M_1})], which exists for $b>1/2$ and $z>1$.
As discussed above, CSR induces a scale-free distribution of $\tij$ [Eq.~(\ref{eq:p(tij)})], ASR and APA together result in a scale-free distribution of $\dti$ [Eq.~(\ref{eq:p(dti)})], and LPA gives a scale-free distribution of $\dtij$ [Eqs.~(\ref{eq:p(dtij,t,t')}) and (\ref{eq:p(dtij,t)})].

\section{Epidemic spreading}
\label{sec:epidemics}
We here give details pertaining to simulations of SI processes on the networks and subsequent analysis of their propagation.

Section~A details how simulations of SI processes were performed and how arrival activity times were calculated.
Section~B defines the symmetrized Kullback-Leibler distance and describes how it was used in practice to compare two distributions of arrival activity times.

\subsection{Simulation of the SI process}
We simulated the SI process as follows~\cite{Panisson2012,Gauvin2013}.

We chose 200 different seeding times $t_0$ (1000 for empirical data) for the epidemic, distributed over the duration of the simulation/measurement.
The interval corresponding to the first 10\% of contacts was excluded to avoid artifacts due to highly transitory behavior for simulated data or related to the handout of RFID badges for empirical data; the last 10\% of the contacts were also excluded to avoid bias due to exclusion of infections that spread slowly.
The seeding times $t_0$ were chosen with probability proportional to the instantaneous activity at $t_0$, $M_1(t_0)$.
For each $t_0$, we initiated a SI process at each node that was active both before and after $t_0$; we recorded the arrival times $t_a$ of the infection at the other nodes of the network.

The arrival activity time $t^*$ of the infection at a node is defined as the number of contacts the node has partaken in during the time-interval $[t_0',t_a]$, where $t_0'$ is the time of the seed's first contact~\cite{Gauvin2013}. (Since the seed may stay isolated for a long time before engaging in contact we use $t_0'$ rather than $t_0$.)

\subsection{Kullback-Leibler divergences}
\label{sec:KL}
Distributions are compared quantitatively by calculating the symmetrized Kullback-Leibler (KL) divergence between them.  The KL divergence between distributions $D^{(1)}=(D^{(1)}_i)_i$ and $D^{(2)}=(D^{(2)}_i)_i$ is defined as~\cite{Kullback1951}:
\beqa
  \KL &=& \frac{1}{2}\sum_{i}D^{(1)}_i\log(D^{(1)}_i/D^{(2)}_i) \label{eq:KL}\\
  &&\quad + \frac{1}{2}\sum_{i}D^{(2)}_i\log(D^{(2)}_i/D^{(1)}_i)  \enspace.
  \nonumber
\eeqa
The KL divergence is only defined for strictly positive $D^{(1)}$ and $D^{(2)}$; in practice, since we calculated $D^{(1)}$ and $D^{(2)}$ from numerical simulations, they may have zeros. To avoid numerical problems when we calculate the KL divergence between models and data, we thus replaced eventual zeros in $p(t^*)$ by $\epsilon$, where we chose $\epsilon$ to be $0.01$ times the smallest value of $p(t^*)$.

The KL divergence is only defined for properly normalized distributions, i.e., distributions that sum to one. To calculate KL10+ we thus first normalized distributions as $D^{(i)}_{\KL10+}=(D^{(i)}_i)_{i=10\ldots}/\sum_{i=10}^{\ldots} D^{(i)}_i$ before calculating the divergences from Eq.~(\ref{eq:KL}).

%\newpage
%=== References: =====================
%\bibliographystyle{h-physrev} %{naturemag}
%\bibliography{References}

%\newpage
\titlepage
\setcounter{page}{12}
\numberwithin{figure}{section}
\renewcommand{\thefigure}{\thesection\arabic{figure}}
\section{Supplementary figures}
\label{sec:supfigs}

\begin{figure*}[h!]
  \includegraphics[width=\textwidth]{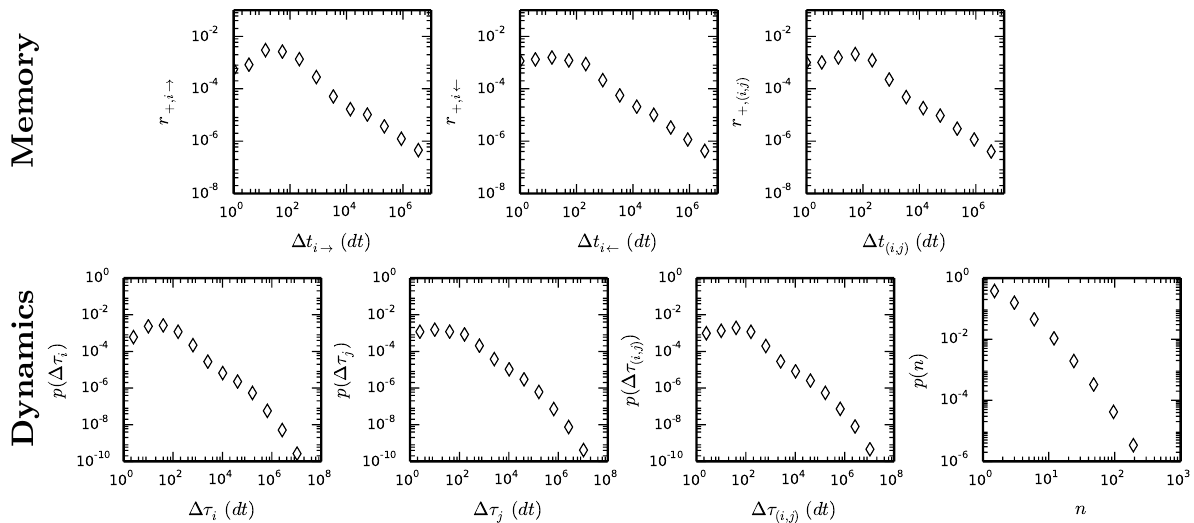}
  \caption{Memory effects and contact dynamics of an empirical network of electronic messages sent between 1,899 students at University of California Irvine (UCI) recorded over 194 days~\cite{Panzarasa2009}. Since the message network is directed, we can here investigate the shape of $\fa$ and $\Pa$ individually.
  a) Average rate $r_{+,i\rightarrow}$ with which an individual $i$ sends a new email as function of the time $\Delta t_{i\rightarrow}$ elapsed since he last send an email. %; $\Delta n_0(\tau_i)/n_0(\tau_i)$ estimates $b\,\fa(\tau_i)$.
  b) Average rate $r_{+,i\leftarrow}$ with which an individual $i$ receives a new email as function of the time $\Delta t_{i\leftarrow}$ elapsed since he last received an email. %; $\Delta n_0(\tau_j)/n_0(\tau_j)$ estimates $r_+\,\Pa(\tau_j)$.
  c) Average rate $r_{+,(i,j)}$ with which an email is send between two individuals $i$ and $j$ (either $i\to j$ or $j\to i$) as function of the time $\Delta t_{(i,j)}$ elapsed since they last exchanged an email. %; $\Delta m_0(\tau)/m_0(\tau)$ estimates $\alpha\,r_+(\tau)\Pl(\tau)$.
  d) Distribution of times $\Delta\tau_i$ elapsed between the times at which an individual $i$ sends two consecutive emails.
  e) Distribution of times $\Delta\tau_j$ elapsed between the times at which an individual $j$ receives two consecutive emails.
  f) Distribution of times $\Delta\tau_{(i,j)}$ elapsed between the times at which two successive emails are sent between two individuals $i$ and $j$.
  g) Distribution of numbers $n$ of mails sent between each pair of individuals $(i,j)$.}
  \label{supfig:memory}
\end{figure*}

\begin{figure*}
  \centering
  \includegraphics[width=\textwidth]{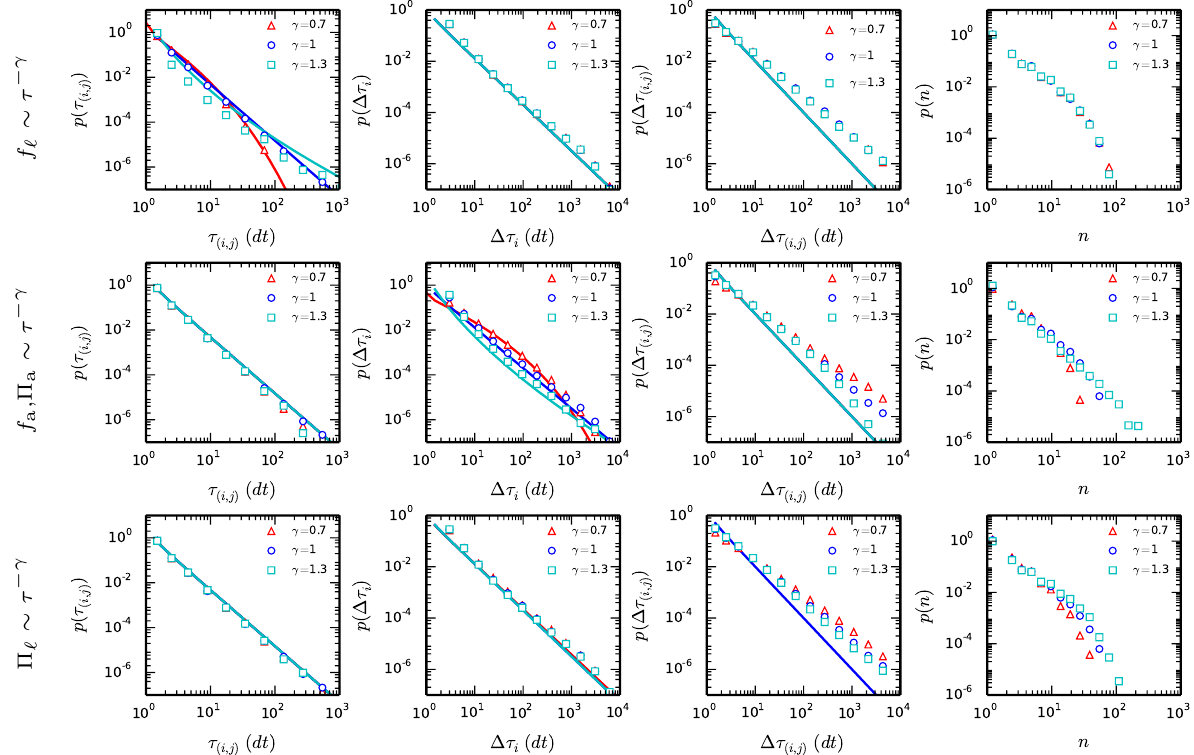}
  \caption{Dynamics of model networks for different functional form of the memory kernels: distributions of contact durations $\tij$, agents' inter-contact durations $\dti$, links' inter-contact durations $\dtij$, and numbers $n$ of contacts per link.
  (top) For $\fl(\tau)=(1+\tau)^{-\gamma}$ and $\Pa(\tau)=\Pl(\tau)=\fa(\tau)=(1+\tau)^{-1}$;
  (middle) For $\Pa(\tau)=\fa(\tau)=(1+\tau)^{-\gamma}$ and $\Pl(\tau)=\fl(\tau)=(1+\tau)^{-1}$;
  (bottom) For $\Pl(\tau)=(1+\tau)^{-\gamma}$ and $\Pa(\tau)=\fl(\tau)=\fa(\tau)=(1+\tau)^{-1}$.
  The number of agents in the network was $N=399$, and $b$ and $z$ were chosen such that the total number of contacts and the total contact duration were approximately the same as in the empirical conf. network.
Symbols correspond to numerics and lines to analytical computations.
  For $\Pl\sim\tau^{-\gamma}$ with $\gamma\neq1$, we do not have an analytical form for $p(\dtij)$.}
  \label{supfig:exponents}
\end{figure*}

\begin{figure*}
  \includegraphics[width=\textwidth]{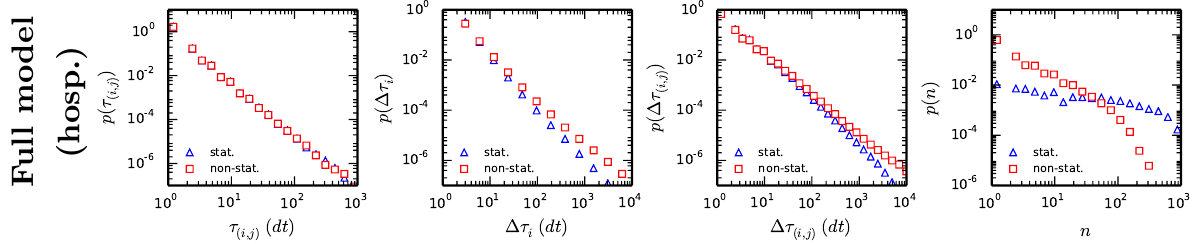}
 \caption{Dynamics of model network with all four memory mechanisms (Full model) for parameters corresponding to the hosp. data. Parameter values are: $N=80$ and $T=17382$, and $b=0.39$ and $z=1.39$ for non-stat., while $b=0.5$ and $z=1.39$ for stat..}
 \label{supfig:hosp}
\end{figure*}

\begin{figure*}
  \includegraphics[width=\textwidth]{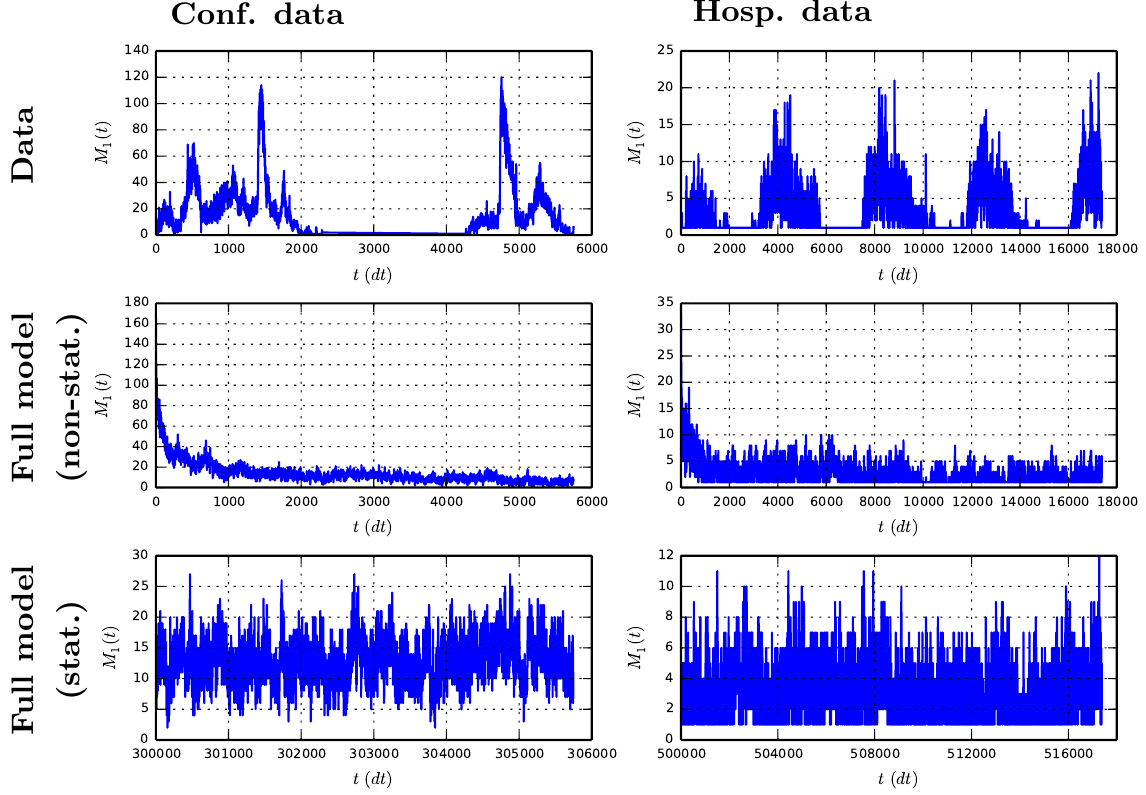}
    \caption{Longitudinal activity $M_1(t)$ for empirical and model networks with all four memory mechanisms. Model parameters were chosen such that the number of agents, total number of contacts, and total contact duration were the same as in the corresponding data (columns).}
    \label{supfig:activity}
\end{figure*}

\begin{figure*}
  \includegraphics[width=\textwidth]{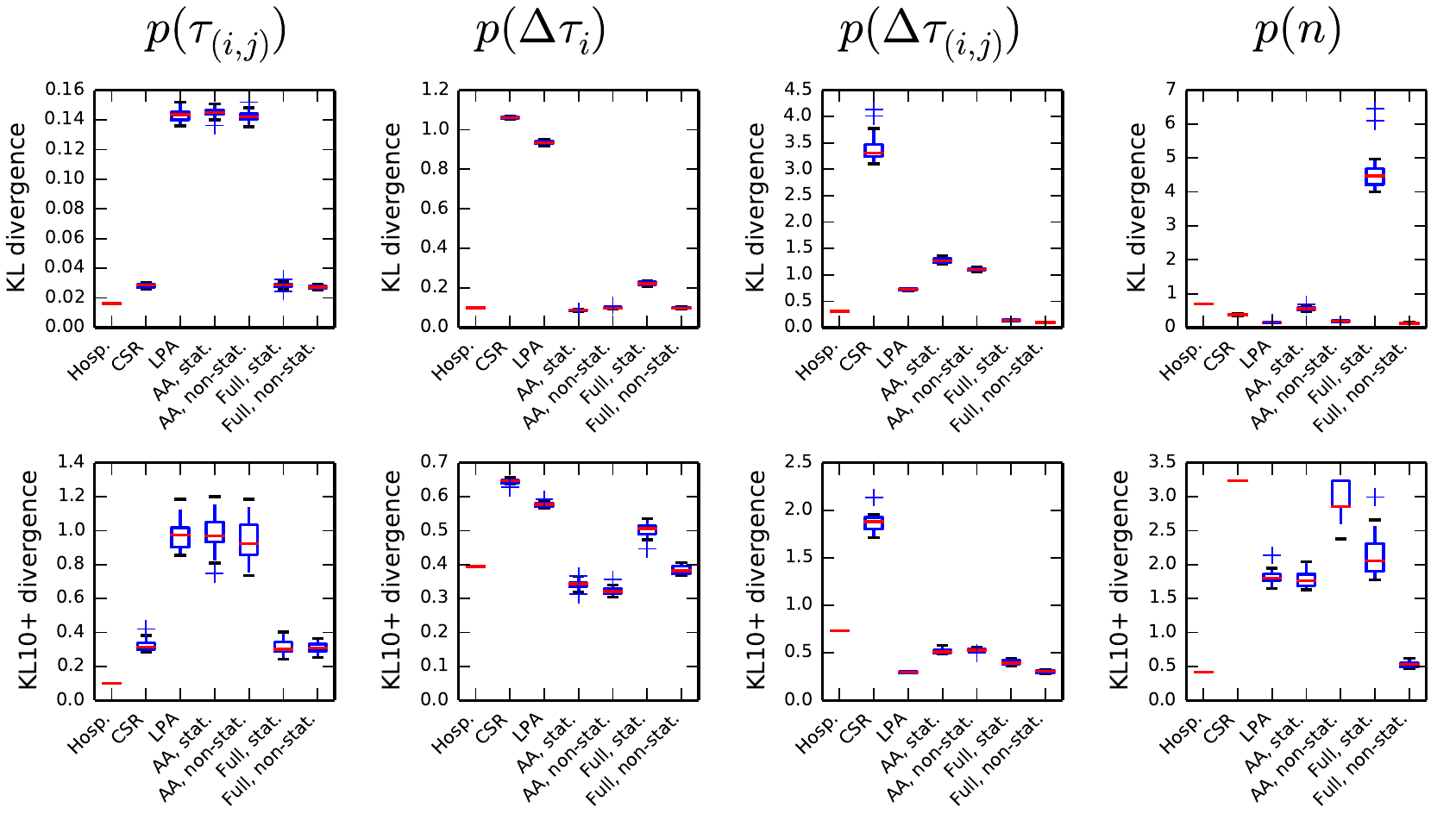}
  \caption{Quantitative comparison of distributions of $\tij$, $\dti$, $\dtij$, and $n$ between model networks and the conf. network: boxplots of symmetrized Kullback-Leibler (KL) divergences (Appendix~\ref{sec:KL}). KL10+ are KL divergences restricted to the distributions for durations or number of contacts larger than 10 $dt$ or 10 for distributions of durations and number of contacts per link, respectively. Note that since distributions are highly skewed, the KL divergence almost exclusively weights low values of $\tau$, $\Delta\tau$, and $n$. To get a more complete picture of the agreement between model and data one should also look at KL10+ which compares the tails of the distributions. KL(10+) divergences between hosp. and conf. networks are also shown for comparison.}
  \label{supfig:KL_dist}
\end{figure*}

\begin{figure*}
  \includegraphics[width=\textwidth]{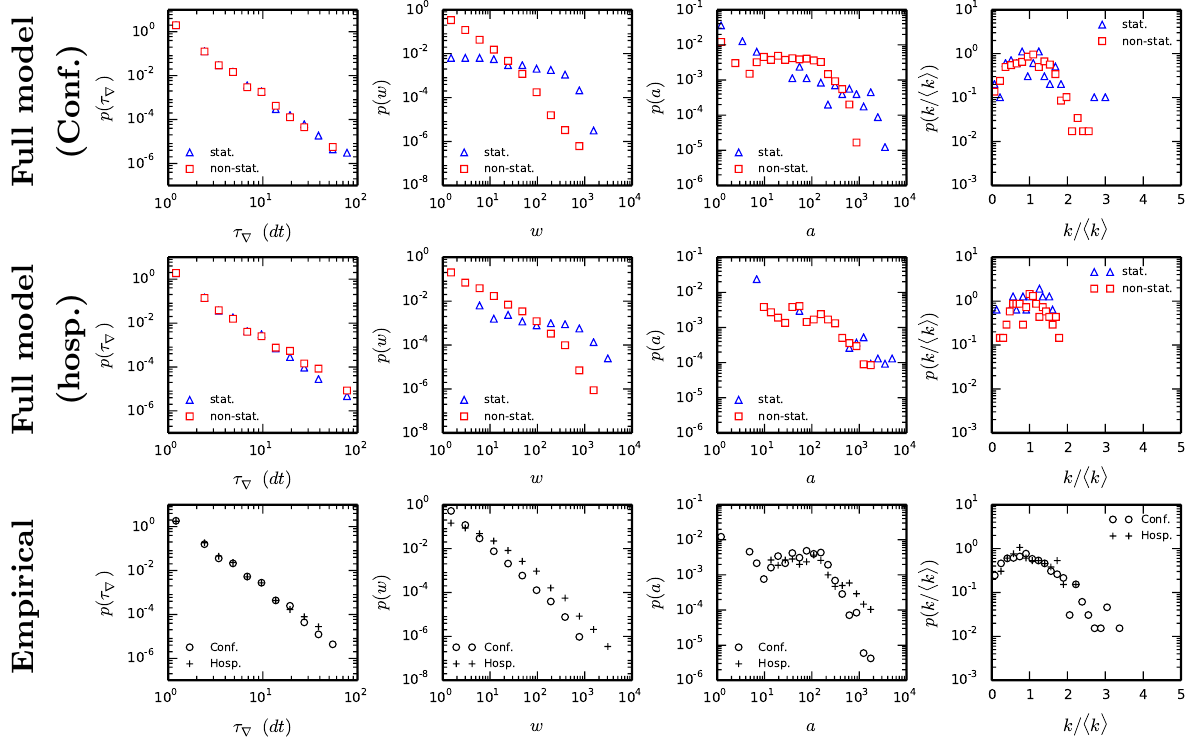}
    \caption{Additional statistics of model networks with all four memory mechanisms (Full model) and empirical networks. Distributions of triangle lifetimes $\tau_\nabla$, of edge weights $w$, of agent activities $a$ (number of contacts an agent partakes in), and of normalized degrees $k/\Ein{k}$. $p(a)$, $p(w)$, and $p(k/\Ein{k})$ are for networks aggregated over the duration of simulations/measurements.
    %See Table~\ref{suptab:stats} for summary statistics for model and empirical networks.
    }
    %Model networks with parameters corresponding to the conf. data showed average clustering coefficients of $0.178\pm0.004$ for non-stationary (non-stat.) and $0.24\pm0.04$ for stationary (stat.) networks compared to $0.121\pm0.001$ and $0.0018\mp0.0005$, respectively, for random (ER) graphs with the same number of nodes and links. The average clustering coefficient of the conf. network was 0.32 compared to $0.16$ for a corresponding random graph.
    %Model networks corresponding to hosp. data showed average clustering coefficients of $0.48\pm0.02$ for non-stat. and $0.75\pm0.07$ for stat. networks compared to $0.370\pm0.007$ and $0.010\mp0.003$, respectively, for random graphs, while the conf. network showed an average clustering coefficient of 0.69 compared to $0.44$ for a corresponding random graph.
    %Note that clustering coefficients from different networks are not directly comparable since the number of links differ ($12265$ for the conf. network and $9600\pm300$ for non-stat. and $340\pm40$ for stat. corresponding model networks; $1405$ for the hosp. network and $1160\pm60$ for non-stat. and $67\pm17$ for stat. model networks);
    %for stationary networks, the number of active nodes ($73\pm6$ for conf. and $16\pm3$ for hosp.) are also different than from  empirical and non-stationary model networks (396-399 for conf. and 79-80 for hosp.).}
    \label{supfig:stats}
\end{figure*}

\begin{figure*}[hb!]
  \includegraphics[width=\textwidth]{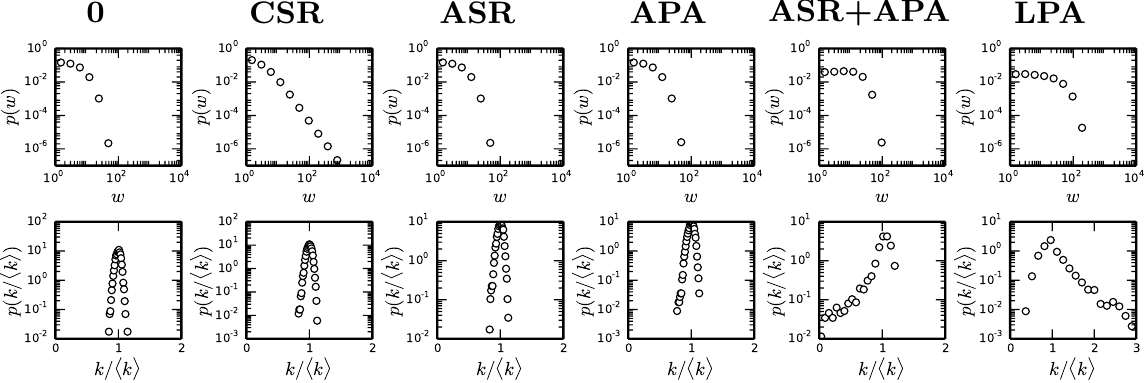}
  \caption{Effect of individual memory mechanisms on topology: distributions of edge weights $w$ and normalized degrees $k/\Ein{k}$ in networks integrated over time-windows of $\Delta T=10$,000.
  Open circles and full lines mark numerical and analytical results, respectively.
  Titles indicate the memory effects included in each column:
  ({\bf 0}) memoryless network ($\fl=\fa=\Pa=\Pl=1$);
  ({\bf CSR}) contact self-reinforcement [$\fl(\tau)=(1+\tau)^{-1}$];
  ({\bf ASR}) activity self-reinforcement [$\fa(\tau)=(1+\tau)^{-1}$];
  ({\bf APA}) agent-centric preferential attachment [$\Pa(\tau)=(1+\tau)^{-1}$].
  ({\bf ASR+APA}) ASR and APA [$\fa(\tau)=(1+\tau)^{-1}$] and $\Pa(\tau)=(1+\tau)^{-1}$];
  ({\bf LPA}) link-centric preferential attachment [$\Pl(\tau)=(1+\tau)^{-1}$].
  For all simulations $N=100$, $dt=0.1$, and the model was run until a stationary state was reached before
  recording statistics; $z$ and $b$ were chosen such that the average rate for an agent to initiate a
  new contact was $r^*_+=0.05$ and the average rate for an existing contact to end was $r^*_-=0.5$ in the
  quasi-stationary state (as witnessed by the mean of $M_1(t)$ remaining constant over time), hence the average number of active links was $M_1^*=10$.}
  \label{supfig:elements}
\end{figure*}

\begin{figure*}
  \includegraphics[width=\textwidth]{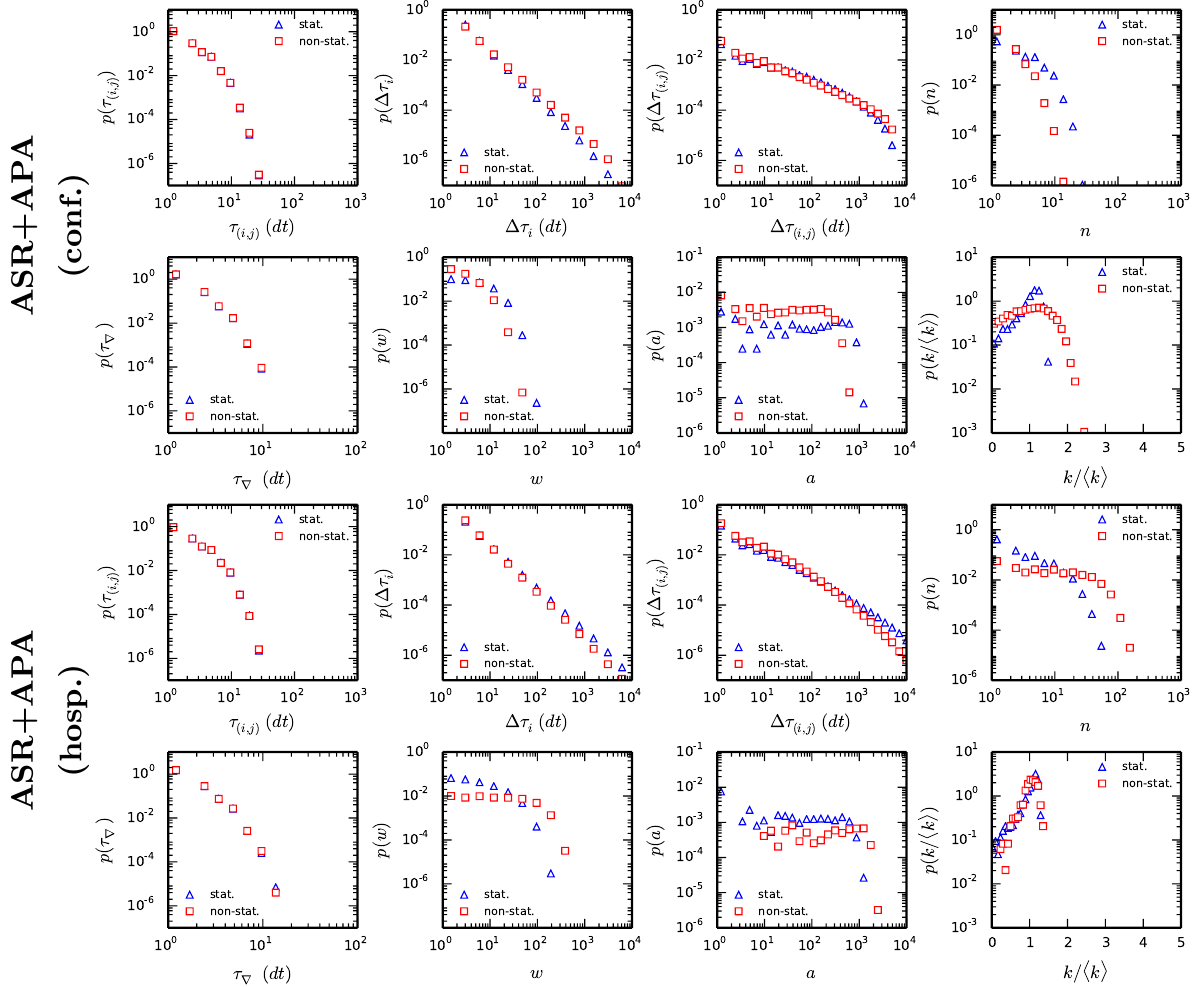}
  \caption{Dynamics of model network with ASR+APA for parameters corresponding to the conference (conf.) data and for parameters corresponding to the hospital (hosp.) data. Parameter values are: for conf., $N=399$ and $T=5750$, with $b=0.314$, $z=0.44$ for non-stat. and $b=0.4$, $z=0.44$ for stat.; for hosp., $N=80$ and $T=17382$, with $b=0.3$, $z=0.39$ for non-stat. and $b=0.35$, $z=0.39$ for stat..
  %See Table~\ref{suptab:stats} for summary statistics for model and empirical networks.
  }
  %Average clustering coefficients for model networks corresponding to conf. data were $0.441\pm0.008$ for non-stationary (non-stat.) and $0.83\pm0.01$ for stationary (stat.) systems compared to $0.285\pm0.002$ and $0.691\pm0.007$ for random (ER) networks with the same number of nodes and links.
  %Average clustering coefficients for model networks corresponding to hosp. data were $0.84\pm0.02$ for non-stat. and $0.92\pm0.02$ for stat. systems compared to $0.748\pm0.008$ and $0.182\pm0.008$ for random networks. ASR+APA thus induce significantly higher clustering in the integrated networks compared to random networks. Note, however, that the clustering cannot be directly compared to empirical or model networks containing all four memory mechanisms since the number of links differ. Integrated networks produced by the model with ASR+APA are significantly more dense ($22500\pm700$ and $2400\pm100$ links for non-stat. networks corresponding to conf. and hosp. data, respectively) than full model and empirical networks ($12265$ and $1405$ for conf. and hosp. networks and $9600\pm300$ and $1160\pm60$ for non-stat. full model networks).}
  \label{supfig:asr+apa}
\end{figure*}

\begin{figure*}
  \includegraphics[width=\textwidth]{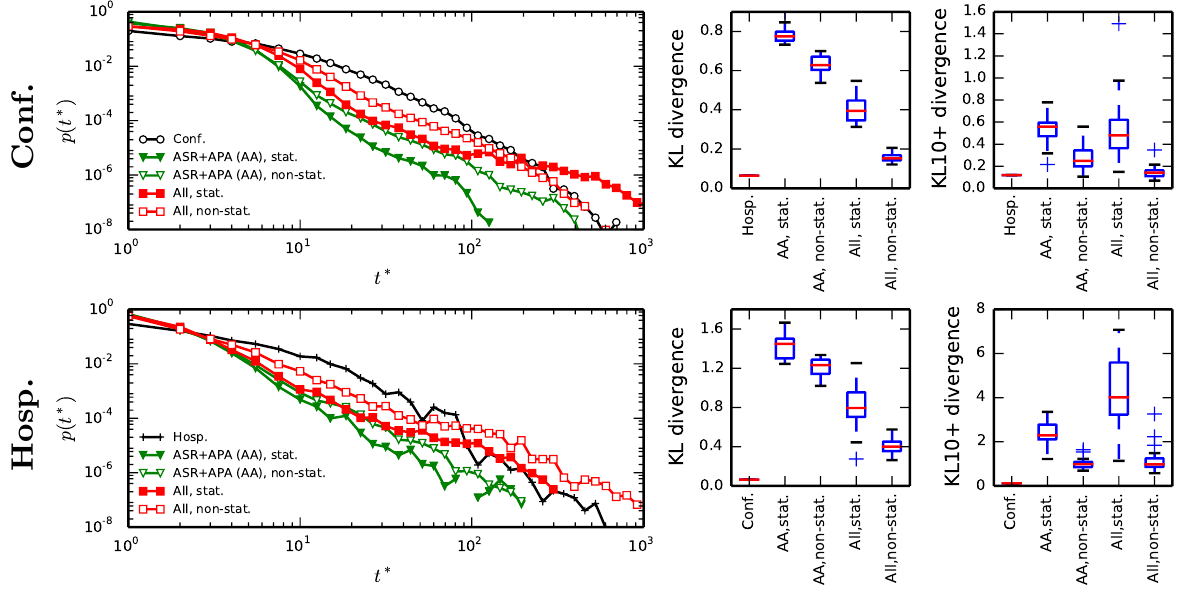}
  \caption{(left) Arrival time$^*$ distributions $p(t^*)$ of SI processes on model networks with parameters corresponding to conf. (top) and hosp. data.
  (bottom) and (right) boxplots of the corresponding KL and KL10+ divergences.}
  \label{supfig:SI}
\end{figure*}

%============================================================================
\end{document}